\documentclass[a4paper,11pt]{article}

\pdfoutput=1

\usepackage[utf8]{inputenc}
\usepackage{amsmath}
\usepackage{amssymb}
\usepackage{amsfonts}
\usepackage{mathrsfs}
\usepackage{graphicx}
\usepackage{booktabs,adjustbox}
\usepackage{caption}
\usepackage{slashed}
\usepackage{verbatim}
\usepackage{float}
\usepackage{setspace}
\usepackage{subfig}
\usepackage{jheppub}

\usepackage{color}
\usepackage{ulem}

\usepackage{bookmark}
\usepackage{wasysym}

\makeatletter
\newcommand{\thickhline}{%
	\noalign {\ifnum 0=`}\fi \hrule height 1pt
	\futurelet \reserved@a \@xhline
}
\makeatother

\usepackage[utf8]{inputenc}
\usepackage{float}

\allowdisplaybreaks[0]

\title{\boldmath Observability of CMB spectrum distortions  from   dark matter    annihilation  }

\author[a]{Shao-Ping Li}

\affiliation[a]{Institute of High Energy Physics, Chinese Academy of Sciences, Beijing 100049, China}
\emailAdd{spli@ihep.ac.cn}

\abstract{Even after  dark matter chemically freezes out in the early universe, electromagnetic cascades from  dark matter annihilation can still perturb the background photon spectrum when the universe temperature cools down to 0.5~keV. We   revisit the    CMB   spectrum distortions  caused by $s$-wave dark matter annihilation under the updated Planck data and the future CMB sensitivity,  concluding  that $s$-wave annihilation cannot create observable  distortions  under forecast sensitivities of the (Super-)PIXIE missions. We further detail the case of $p$-wave   dark matter annihilation, demonstrating  the observability of   the primordial  $\mu$-distortion. Taking  current constraints from primordial light elements, structure formations,  cosmic electron-positron rays, and gamma rays, we find that the $\mu$-distortion reaching the observational limit as large as $\mu\simeq 3\times 10^{-8}$  can only be realized with a dark matter mass at 10--50~MeV and  a kinetic decoupling temperature around 1~keV.  The upper bound of the  $p$-wave annihilation  cross section can be strengthened by  an order of magnitude    if the $\mu$-distortion is not detected. 
	
}

\begin{document}
	\maketitle
	\flushbottom
	
	\section{Introduction}
	\label{sec:intro}
Over the past decades,  observations of cosmic microwave background (CMB), such as the temperature and polarization anisotropies, have  not only made the cosmology a precise science,  but also provided the most stringent tests and constraints for  the standard $\Lambda$CDM model and theories beyond. Yet, the cosmology will not end up with the CMB anisotropies only.  The CMB possesses another invaluable treasure: the energy (frequency) spectrum.  Measurements of the COBE/FIRAS  experiments~\cite{Mather:1993ij,Fixsen:1996nj} have convinced us that the average CMB spectrum is perfectly close to a  blackbody with a  temperature $T_0=2.73$~K at present day~\cite{Fixsen_2009}.

Any departure  of the CMB  spectrum from a perfect blackbody is referred to as spectral distortions, which can encode the remnant created in the early universe  up to a   temperature of 0.5~keV, or to a redshift as high as $z\sim 10^6$. There are two widely considered distortions, called $\mu$ and $y$ distortions. The former is created typically at redshifts $5\times 10^4\lesssim z\lesssim 2\times 10^6$ and the latter formed below $5\times 10^4$~\cite{Burigana:1991eub,Hu:1992dc}, though richer distortions are   expected  during the transition from the $\mu$-era to the $y$-era~\cite{Chluba:2011hw,Khatri:2012tw,Acharya:2018iwh}.    Unlike  CMB experiments that  have  been  improved dramatically  in sensitivities of temperature and polarization anisotropies,  CMB spectral distortion measurements are still dominated by COBE/FIRAS    nearly three decades ago, with the   $95\%$ C.L. upper bounds:
$|\mu|<9\times 10^{-5}, |y|<1.5\times 10^{-5}$.
Such a situation has triggered  several programs that aim to increase the sensitivity of spectrum distortions by at least three orders of magnitude. Known projects include PIXIE~\cite{Kogut:2011xw}, PIXIE-like (Super-PIXIE\footnote{We will hereafter refer to this program as  Super-PIXIE following   Ref.~\cite{Chluba:2019nxa}.})~\cite{Kogut:2019vqh}, PRISM~\cite{PRISM:2013fvg}	and Voyage 2050 programs~\cite{Chluba:2019nxa}. For example, the (Super-)PIXIE missions can   reach  sensitivities of  $|\mu|\simeq (0.8-3) \times 10^{-8},  |y|\simeq (2-3) \times 10^{-9}$.   
In particular,  the raw sensitivity of the PIXIE mission has the $5\sigma$ detection forecast for $|\mu|=5\times 10^{-8}$ and $|y|=1\times 10^{-8}$~\cite{Kogut:2011xw}.

Increasing the detection sensitivity of the CMB spectrum distortions has multi-fold significance. Within the standard $\Lambda$CDM model, there are sources that can create CMB spectrum distortions. The largest effects are expected from the damping of primordial small-scale perturbations, with  $\mu\simeq 2\times  10^{-8}$ possibly inherited from inflation~\cite{Cabass:2016giw,Chluba:2016bvg}, and  from  reionization and structure formation, with $y\sim 10^{-6}$ created at $z\lesssim 10$ due to the Sunyaev-Zeldovich effects~\cite{Refregier:2000xz,Dolag:2015dta,Hill:2015tqa,DeZotti:2015awh}.
Therefore,  increasing the detection sensitivity   of CMB spectrum distortions   will provide another independent and complementary test of the standard $\Lambda$CDM model. 

Another source  which could  create significant distortions is the cold dark matter (DM) in the $\Lambda$CDM model. While all the observations of DM thus far are related to gravitational effects, the origin of the DM relic density is not yet  determined experimentally.  
With minimal assumptions of DM couplings to the Standard Model (SM) particles,  the relic density can   be predicted  to match the observed value. Once the interaction  between DM and SM particles is  introduced to explain the relic density,  it could then create the CMB spectrum distortions via residual DM annihilation to SM particles even after the DM abundance is frozen in the early universe.\footnote{For not absolutely stable DM, DM decay could also generate CMB spectrum distortions.} Therefore,
  CMB spectrum distortions  can  also provide  a promising avenue to study the nature of DM, particularly the DM mass scale and  interactions with SM particles beyond the universal gravity. 
  
CMB spectrum distortions from DM annihilation have been considered for a while in  the  so-called $s$- and $p$-wave annihilation channels. The $s$-wave channel assumes that the   cross section is dominated by DM velocity-independent  annihilation while the $p$-wave annihilation  channel depends on the DM squared velocity.  
For $s$-wave DM annihilation, however, it has been noticed that generating large distortions  is strongly limited, since    energy injection would also cause large impacts at later recombination and reionization epochs~\cite{Padmanabhan:2005es,Chluba:2009uv,Chluba:2011hw,Chluba:2013wsa}.  Some attention was also paid to   $p$-wave DM annihilation, where CMB constraints are much weaker due to  the velocity-dependent  annihilation cross section~\cite{McDonald:2000bk,Chluba:2013wsa,Chluba:2013pya}. 
Nevertheless, even if the   robust  constraints from CMB anisotropies are weaker by orders of magnitude,   large  $p$-wave annihilation to SM particles could still  be limited   by big-bang nucleosynthesis (BBN) and astrophysical observations. 

Given   the available limitations from BBN and astrophysical measurements, however,
 it is currently not  clear yet to what extent   we can expect observable CMB spectrum distortions from future missions, and  particularly what   we can learn from the distortion signals about  the DM mass scale and particle interactions. 
  In this paper, we take a step forward to investigate the CMB distortions from simple DM annihilation by   applying  the limitations  from CMB anisotropies, measurements of primordial light elements, structure formations, cosmic $e^\pm$ rays, and $\gamma$ rays.  It proves to be of benefit and significance when we make such a   synergy, since it allows us to check whether the distortion signals, if observed,  are sourced by DM annihilation. On the other hand, if no DM-sourced signals, after proper extraction,   are detected in the future, CMB spectrum distortions will provide competitive  and even stronger limits on   DM mass and interactions. 

 We begin in Sec.~\ref{sec:pri-dist} with the CMB spectrum distortions from general energy injection. Then in Sec.~\ref{sec:DMann}, we make  minimal assumptions to consider  the CMB distortions from   DM $s$- and $p$-wave    annihilation without specifying the particle physics models,  and present the observable regimes under the measurements of CMB,  BBN and astrophysics in Sec.~\ref{sec:detection}, demonstrating the challenges and possibilities of observable distortions from the two  typical  DM annihilation modes.  Conclusions are drawn in Sec.~\ref{sec:con} and some technical details are relegated to appendices.  In this paper, natural units are used: $\hbar=k_B=c=1$.

\section{CMB spectrum distortions from energy injection}\label{sec:pri-dist}
\subsection{The $\mu$-era}
At high redshift, $z\gtrsim z_{\mu}\approx 2\times 10^6$~\cite{Burigana:1991eub,Hu:1992dc}, or the temperature of the universe is higher than about 0.5~keV, the rapid chemical thermalization processes from double Compton scattering ($\gamma+e^-\to e^-+\gamma +\gamma$) and bremsstrahlung ($e^- +N\to e^-+N+\gamma$), as well as the kinetic thermalization process from Compton scattering ($\gamma+e^-\to \gamma+e^-$) keep the photon distribution in the perfect blackbody spectrum, $f_\gamma=(e^{x}-1)^{-1}$, with $x\equiv E_\gamma/T$ and $T$ the background temperature. When   $z\lesssim z_{\mu}$, the chemical thermalization processes become inefficient for high-frequency photons while   kinetic thermalization is still faster than the Hubble expansion until $z\lesssim z_{\mu y} \approx 5\times 10^{4}$~\cite{Hu:1992dc}. During $z_{\mu y}\lesssim z\lesssim z_{\mu}$, or $10~\text{eV}\lesssim T\lesssim 500~\text{eV}$, the photon distribution is characterized by the Bose-Einstein statistics with a non-vanishing  and   frequency-dependent chemical potential $\mu(x)$. 

We can write  the photon distribution with a  $\mu$-distortion as
\begin{align}\label{eq:fmu-dis}
	f_\gamma(x_e,\mu)=\frac{1}{e^{x_e+\mu}-1}
	\approx \frac{1}{e^{x_e}-1}-\mu \frac{e^{x_e}}{(e^{x_e}-1)^2}\,,
\end{align}
where $\mu$ is normalized to temperature, with $x_e\equiv E_\gamma/T_e$. Here, we have chosen  the electron temperature   $T_e$ as the reference temperature. Note that while  the photon-electron plasma can establish a  nearly identical temperature via Compton/inverse Compton scattering down to the recombination epoch,  $T_e$ is, due to energy injection,  in general different from the background temperature $T=T_0(1+z)$, where $T_0=2.73~\text{K}\approx 2.35\times 10^{-4}$~eV is the present CMB temperature.  Nevertheless, the temperature difference is expected to be at the same order of primordial distortions.\footnote{When we discuss primordial distortions, we are referring to the $\mu$ and $y$ distortions formed at the pre-recombination epoch, $z>z_{\rm rec}\simeq 1100$.}    In the following, we will set the photon temperature as the background temperature,  $T_\gamma\approx T$, but distinguish the difference between $T$ and $T_e$ whenever necessary. More details on the temperature difference are presented in  Appendix~\ref{app:Tdif}.

To evaluate the evolution of $\mu$, one can start from the Boltzmann equation:
\begin{align}\label{eq:f-Boltzmann}
	\frac{\partial f_\gamma}{\partial t}-H p\frac{\partial f_\gamma}{\partial p}=\frac{\delta f}{\delta t}\Big|_{\rm dC}+\frac{\delta f}{\delta t}\Big|_{\rm brem}+\frac{\delta f}{\delta t}\Big|_{\rm C}+\frac{\delta f}{\delta t}\Big|_{\rm inj}\,,
\end{align}
where $H=H(T)$ is the Hubble parameter with the background temperature $T$,  and the collision rates correspond to double Compton scattering, bremsstrahlung, Compton scattering, and exotic energy injection, respectively.   
The rates (dC, brem, C) and their characteristic time scales are collected in Appendix~\ref{app:rates} for reference. The energy injection process can   come  either from direct photon injection or from electron-positron injection, the small difference of which   is discussed in Appendix~\ref{app:e-channel}.
When a small amount of   exotic  energy is injected to the photon plasma, it  is quickly redistributed via Compton scattering (also called \textit{Comptonization}).  The photon-number changing process (dC+brem) is     efficient only at the low-$E_\gamma$ region and exponentially suppressed at the high-$E_\gamma$ band (see also Appendix~\ref{app:rates}), indicating an $E_\gamma$-dependent chemical potential. However, given the photon energy band  covered by the future mission programs, like the (Super-)PIXIE and PRISM experiments, is  $E_\gamma=[30, 6000]$~GHz,\footnote{It corresponds to $x_0=E_\gamma/T_0\approx [0.5,105.6]$. $1$~GHz $\approx 4.14\times 10^{-6}$~eV.} we will  neglect the frequency dependence of $\mu$.\footnote{The solution of $\mu$ in the limit $x_e\ll 1$ can be approximated as $\mu(x_e)\approx \mu_c e^{-x_c/x_e}$, with $\mu_c$ a constant distortion formed at large $x_e$ and $x_c\lesssim  0.01$~\cite{ZS1970}. It implies that for future mission programs covering $x_e>0.5$, the $\mu$-distortion is essentially constant.} After Comptonization,
 the approximate equations of photon energy and number densities  at high-$x_e$   thus become
\begin{align}
\frac{\partial \rho_\gamma(x_e,\mu)}{\partial t}+4H \rho_\gamma(x_e,\mu)\approx\frac{\delta \rho}{\delta t}\Big|_{\rm in}\,,\label{eq:rho-Boltzmann-mu}
\\[0.2cm]
\frac{\partial n_\gamma(x_e,\mu)}{\partial t}+3H n_\gamma(x_e,\mu)\approx\frac{\delta n}{\delta t}\Big|_{\rm in}\,.\label{eq:n-Boltzmann-mu}
\end{align}
Here the photon energy and number densities at    leading order  are given by
\begin{align}\label{eq:rho-n_gamma}
	\rho_\gamma(x_e,\mu)=\rho_\gamma(x_e)(1-a_\mu\mu)\,, \quad n_\gamma(x_e,\mu)=n_\gamma(x_e)(1-b_\mu\mu)\,,
\end{align}
 where $\rho_\gamma(x_e), n_\gamma(x_e)$ follow the blackbody distribution, and 
 \begin{align}
 	a_\mu&=\frac{\int dx x_e^3 e^{x_e}/(e^{x_e}-1)^2}{\int dx x_e^3 /(e^{x_e}-1)}\approx 1.11\,,
 	\\[0.2cm]
 		b_\mu&=\frac{\int dx x_e^2 e^{x_e}/(e^{x_e}-1)^2}{\int dx x_e^2/(e^{x_e}-1)}\approx 1.37\,.
 \end{align}
In Eqs.~\eqref{eq:rho-Boltzmann-mu}-\eqref{eq:n-Boltzmann-mu}, there are two kinds of small perturbations, the chemical potential $\mu$ and  the temperature difference between $T$ and $T_e$. The latter appears as $d\ln(T/T_e)/d\ln T$, and    cannot simply be omitted  when keeping the $\mu$  perturbation at  leading order, since the temperature perturbation and the  $\mu$-distortion can be at the same order.   However, the dependence on $d\ln(T/T_e)/d\ln T$  can be removed by combining Eq.~\eqref{eq:rho-Boltzmann-mu} and Eq.~\eqref{eq:n-Boltzmann-mu}, leading to the equation
\begin{align}\label{eq:dmudt}
	\frac{\partial \mu}{\partial t}=\frac{1}{4b_\mu-3a_\mu}\left(3\frac{1}{\rho_\gamma}\frac{\delta \rho }{\delta t}\Big|_{\rm inj}-4\frac{1}{n_\gamma}\frac{\delta n}{\delta t}\Big|_{\rm inj}\right).
\end{align}
Integrating over the time period during which the $\mu$-distortion can be formed, we arrive at
\begin{align}\label{eq:mu-dist}
	\mu\approx1.4 \frac{\delta \rho }{\rho_\gamma}\Big|_{\rm inj}-1.9 \frac{\delta n}{n_\gamma}\Big|_{\rm inj}\,,
\end{align}
with $\delta \rho/\rho_\gamma, \delta n/n_\gamma$ the integrated collision rates per unit photon energy/number density:
\begin{align}
 \frac{\delta \rho }{\rho_\gamma}\Big|_{\rm inj}&\equiv -\int^\infty_0 \mathcal{J}_\mu(z) \frac{1}{\rho_\gamma}\frac{\delta \rho }{\delta t}\frac{dt}{dz}dz\,,
 \\[0.2cm]
  \frac{\delta n }{n_\gamma}\Big|_{\rm inj}&\equiv -\int^\infty_0 \mathcal{J}_\mu(z) \frac{1}{n_\gamma}\frac{\delta n }{\delta t}\frac{dt}{dz}dz\,.
\end{align}
The  redshift-dependent function, or the visibility function,  $\mathcal{J}_\mu(z)$ encapsulates the epoch during which the primordial $\mu$-distortion can be formed. The approximations for the  visibility function will be discussed in Sec.~\ref{sec:mu-to-y}.

\subsection{The $y$-era}

For $z_{\rm rec}\lesssim z \lesssim z_{\mu y}$, with $z_{\rm rec}\simeq 1100$   the redshift at the recombination epoch,   Compton scattering becomes inefficient to establish a full Bose-Einstein distribution for the photon field, though a partial redistribution of photon energy via Compton scattering is still active. During this epoch, the so-called $y$-distortion is formed.  In general, a dedicated analysis of the CMB distortion at this epoch  requires  numerical solutions of the Boltzmann equation, or of the reduced Kompaneets equation~\cite{Novikov2006}. Nevertheless, approximate analytic solutions exist in two   asymptotic regimes: the high electron temperature limit, $T_e\gg T_\gamma$, and the strong Compton cooling limit. The former is expected with hot electrons   in the intergalactic medium, and  was used to derive the Sunyaev-Zeldovich effect within the $\Lambda$CDM model~\cite{ZS1969}.\footnote{Also see e.g. Ref.~\cite{Novikov2006} Sec.~2 for an updated derivation.}
In the latter regime, the  strong Compton cooling limit  renders the temperature difference between electrons and photons  small, $|T_e-T|/T\ll 1$, which works as a proxy to estimate the primordial $y$-distortion  at $z> z_{\rm rec}$.

To derive the $y$-distortion in the limit of $|T_e-T|/T\ll 1$, we can use the approximation $f_\gamma+f^2_\gamma\approx -T \partial f_\gamma(T)/\partial E_\gamma \approx -(T/T_e)\partial f_\gamma(x_e)/\partial x_e$ in the   last two recoil terms  given in Eq.~\eqref{eq:C}. The Comptonization rate in Eq.~\eqref{eq:C} can then be rewritten as 
 \begin{align}
	\frac{\delta f}{\delta t}\Big|_{\rm C}&= n_e \sigma_T\frac{T_e-T}{m_e}\frac{1}{x_e^2}\frac{\partial }{\partial x_e}\left(x_e^4\frac{\partial f_\gamma}{\partial x_e}\right),\label{eq:C-y}
\end{align}
implying that the $y$-distortion is characterized by the small temperature difference. Introducing  the dimensionless Compton $y_{}$-parameter:
\begin{align}
	\frac{dy}{dt}\equiv   n_e \sigma_T\frac{T_e-T}{m_e}\,,
\end{align} 
and taking the approximation $\partial f_\gamma(x_e,0)/\partial t-Hx_e \partial f_\gamma(x_e,0)/\partial x_e\approx 0$, 
we arrive at a simplified Boltzmann equation:
\begin{align}
\frac{\partial f_y}{\partial t}=\frac{dy}{dt}\frac{1}{x_e^2}\frac{\partial }{\partial x_e}\left(x_e^4\frac{\partial f_\gamma}{\partial x_e}\right),\label{eq:C-y2}
\end{align}
where   the photon distribution is defined as $f_\gamma=f_\gamma(x_e,0)+f_y$ with $f_y\ll f_\gamma(x_e,0)$.
It finally leads to
\begin{align}\label{eq:fy}
f_y\approx y \frac{x_e e^{x_e}}{(e^{x_e}-1)^2}\left(x_e\frac{e^{x_e}+1}{e^{x_e}-1}-4\right)\,,
\end{align}
which is identical to the $T_e\gg T$ limit~\cite{ZS1969}.
It should be emphasized  that the approximation leading to the above equation neglects the   Hubble expansion effect. Consequently,
   the $y$-distortion  is somewhat overestimated, since a portion of photon energy/number density should have been diluted by redshifting.   For   energy injection from DM annihilation, it will be seen in the next section that the   $y$-distortion     is hardly observable by future missions. Besides, it is several orders of magnitude below  the largest  $y$-distortion expected in the   $\Lambda$CDM model.
   Therefore, we will not go into  the precise numerical analysis of  the primordial $y$-distortion from DM annihilation. In   subsequent discussions, we will follow the strong Compton cooling  approximation to estimate the $y$-distortion formed at $z_{\rm rec}\lesssim z \lesssim z_{\mu y}$. 
   
Using Eq.~\eqref{eq:fy}, we can derive the equations for photon energy and number density. 
Similar to Eqs.~\eqref{eq:rho-Boltzmann-mu}-\eqref{eq:n-Boltzmann-mu}, we arrive at
\begin{align}
	\frac{\partial \rho_\gamma(x_e,y)}{\partial t}+4H \rho_\gamma(x_e,y)\approx\frac{\delta \rho}{\delta t}\Big|_{\rm inj}\,,\label{eq:rho-Boltzmann-y}
	\\[0.2cm]
	\frac{\partial n_\gamma(x_e,y)}{\partial t}+3H n_\gamma(x_e,y)\approx\frac{\delta n}{\delta t}\Big|_{\rm inj}\,.\label{eq:n-Boltzmann-y}
\end{align}
Here the photon energy and number densities  at leading order  read
\begin{align}
	\rho_\gamma(x_e,y)=\rho_\gamma(x_e)(1+a_y y)\,, \quad n_\gamma(x_e,y)=n_\gamma(x_e)(1+b_y y)\,,
\end{align}
with  $a_y, b_y$ given by
 \begin{align}
	a_y&=\frac{\int dx_e x_e^4 e^{x_e} [x_e (e^{x_e}+1)/(e^{x_e}-1)-4]/(e^{x_e}-1)^2}{\int dx_e x_e^3 /(e^{x_e}-1)}=4\,,
	\\[0.2cm]
	b_y&=\frac{\int dx_e x_e^3 e^{x_e} [x_e (e^{x_e}+1)/(e^{x_e}-1)-4]/(e^{x_e}-1)^2}{\int dx_e x_e^2/(e^{x_e}-1)}=0\,.
\end{align}
It is noticeable that $b_y=0$ implies at leading order in Eq.~\eqref{eq:n-Boltzmann-y},     particle-number injection only causes  the   temperature difference  while the effect on $\partial y/\partial t$ goes beyond the leading order. However,      the temperature difference also appears in Eq.~\eqref{eq:rho-Boltzmann-y}, so the effect of     particle-number injection can still be transferred to  $\partial y/\partial t$ at leading order. Finally it leads to
  the evolution of the $y$-distortion
\begin{align}
	\frac{\partial y}{\partial t}=\frac{1}{3a_y-4b_y}\left(3\frac{1}{\rho_\gamma}\frac{\delta \rho }{\delta t}\Big|_{\rm inj}-4\frac{1}{n_\gamma}\frac{\delta n}{\delta t}\Big|_{\rm inj}\right).
\end{align}
Integrating over the time period during which the $y$-distortion can be formed, we arrive at
\begin{align}\label{eq:y-dist}
y\approx \frac{1}{4} \frac{\delta \rho }{\rho_\gamma}\Big|_{\rm inj}-\frac{1}{3}\frac{\delta n}{n_\gamma}\Big|_{\rm inj}\,,
\end{align}
with $\delta \rho/\rho, \delta n/n$ the integrated collision rates  per unit photon energy/number density:
\begin{align}
	\frac{\delta \rho }{\rho_\gamma}\Big|_{\rm inj}&\equiv -\int^\infty_0 \mathcal{J}_y(z) \frac{1}{\rho_\gamma}\frac{\delta \rho }{\delta t}\frac{dt}{dz}dz\,,
	\\[0.3cm]
	\frac{\delta n }{n_\gamma}\Big|_{\rm inj}&\equiv -\int^\infty_0 \mathcal{J}_y(z) \frac{1}{n_\gamma}\frac{\delta n }{\delta t}\frac{dt}{dz}dz\,.
\end{align}
Similar to  $\mathcal{J}_\mu(z)$, the  visibility  function $\mathcal{J}_y(z)$ encapsulates the epoch during which the  $y$-distortion can be formed at the pre-recombination epoch.

\subsection{The $\mu$-$y$ transition}\label{sec:mu-to-y}

In the approximation of instantaneous transition between the $\mu$  and $y$ eras at $z=z_{\mu y}\approx 5\times 10^4$ and no large distortions are created at $z\gtrsim z_{\mu}\approx 2\times 10^6$,  $	\mathcal{J}_{\mu, y}(z)$ are given by
\begin{align}\label{eq:visibility-fun}
	\mathcal{J}_\mu(z)= \theta(z_{\mu}-z)\theta(z-z_{\mu y})\,, \quad \mathcal{J}_y(z)=\theta(z_{\mu y}-z)\theta(z-z_{\rm rec})\,, 
\end{align}
where $\theta(z)$ is the Heaviside function.  By relaxing the assumption of no distortions above $z_{\mu}$,  $\mathcal{J}_\mu$  becomes  $\mathcal{J}_\mu\approx e^{-(z/z_\mu)^{5/2}}\theta(z-z_{\mu y})$~\cite{Burigana:1991eub,Hu:1992dc}, which  is exponentially suppressed at $z>z_\mu$. More refined approximations  can be found in Refs.~\cite{Chluba:2013kua,Chluba:2016bvg}.  Given the distortions obtained from these   approximations usually lead to variations at $\mathcal{O}(10\%)$~\cite{Chluba:2016bvg}, we will simply take Eq.~\eqref{eq:visibility-fun} to estimate the primordial distortions generated by DM annihilation.
It should be mentioned that the approximation of Eq.~\eqref{eq:visibility-fun} will not include   distortions created around the transition epoch, typically  at  $1.5\times 10^4\lesssim z\lesssim 2\times 10^5$.\footnote{For example, the  intermediate-type distortion~\cite{Khatri:2012tw} and the  nonthermal relativistic distortion~\cite{Acharya:2018iwh} that can help to distinguish various energy injection sources.} Nevertheless,  for observable   distortions induced by DM annihilation considered in this paper, it turns out that the largest distortion is formed at higher redshift $z>z_{\mu y}$. Given this reason, we will not consider any further distortions formed  at  $1.5\times 10^4\lesssim z\lesssim 2\times 10^5$.


\section{Energy injection from DM annihilation}\label{sec:DMann}

The primordial distortions given in Eqs.~\eqref{eq:mu-dist} and~\eqref{eq:y-dist}  depend on the   energy transfer   and the photon-number generation rates. The photon-number changing process from DM annihilation can come either from direct production or from final-state radiation, e.g., $\text{DM}+\text{DM}\to e^++e^-+\gamma$.  In the former case, let us suppose that   DM dominantly annihilates into photons, with the thermally averaged cross section $\langle\sigma v\rangle_\gamma$. We  then expect
\begin{align}
	\frac{1}{\rho_\gamma }\frac{\delta \rho }{\delta t}\propto \frac{m_{\rm DM}\langle\sigma v\rangle_\gamma n_{\rm DM}^2}{T^4}\,, \quad \frac{1}{n_\gamma }\frac{\delta n }{\delta t}\propto\frac{\langle\sigma v\rangle_\gamma n_{\rm DM}^2}{T^3}\propto \frac{T}{m_{\rm DM}}
	\left(\frac{1}{\rho_\gamma }\frac{\delta \rho }{\delta t}\right)\,.
\end{align}
Here we have used the approximation that the photon energy inherited from each nonrelativistic DM annihilation is typically at $m_{\rm DM}$.
For  thermal DM, the mass below 10~MeV is strongly disfavored by BBN~\cite{Serpico:2004nm,Ho:2012ug,Boehm:2013jpa,Nollett:2013pwa,Nollett:2014lwa,Kawasaki:2015yya,Depta:2019lbe,Sabti:2019mhn}, while for   nonthermal DM, a mass below a few keV is   limited by   Lyman-$\alpha$ observations~\cite{Ballesteros:2020adh,DEramo:2020gpr,Decant:2021mhj}. Given that  the photon temperature during the epoch of primordial distortion formation is smaller than 0.5~keV, the number changing rate is   suppressed compared to the energy-transfer rate. 

On the other hand, if the dominant photon-number changing process stems from final-state radiation,  we expect that abundant photon numbers are generated  in the infrared regime $E_\gamma\to 0$.  For energy injection well before recombination,  spectrum distortions from these low-energy photons  would  be processed by rapid double Compton scattering and bremsstrahlung~\cite{McDonald:2000bk}.  However, when energy injection occurs near or after the recombination epoch, the soft-photon heating effect  can cause significant temperature difference between $T_\gamma$ and $T_e$,  and can also have non-negligible impacts at $z\lesssim \mathcal{O}(10)$~\cite{Cyr:2024vkt}. 
In the following, we will   focus on   energy injection before recombination. In this case,  the  primordial photon distortions  are dominantly   determined by the energy-transfer rate.

\subsection{Annihilation via $s$-wave}\label{sec:mudist-swave}

It was shown in Refs.~\cite{Padmanabhan:2005es,Chluba:2009uv,Chluba:2011hw,Chluba:2013wsa} that large primordial distortions from DM $s$-wave annihilation suffer from strong constraints of CMB observations. Here we revisit   $s$-wave annihilation using the updated Planck data, demonstrating  that CMB bounds disfavor observational  regimes for the primordial distortions  within forecast sensitivities.  Moreover,  we   anticipate  the largest     distortions  allowed by  future CMB constraints, and discuss the difficulty  of probing such primordial distortions. 

Neglecting the small contribution from    photon-number injection, we  can   parameterize the energy-transfer rate from two-body DM annihilation as
 \begin{align}
 	\frac{\delta \rho}{\delta t}&=f_{\rm EM}\langle \sigma v\rangle m_{\rm DM}n_{\rm DM}^2\label{eq:f-param}
  \\[0.2cm]
 	&\approx \left(\frac{8.7\times 10^{-13}}{\text{eV}^{-1}\text{cm}^3~\text{s}}\right) \left(\frac{1+z}{1000}\right)^6\left(\frac{\Omega_{\rm DM}h^2}{0.12}\right)^2\left(\frac{f_{\rm EM}\langle \sigma v\rangle }{6\times 10^{-26}\text{cm}^3/\text{s}}\right)\left(\frac{100~\text{GeV}}{m_{\rm DM}}\right),\label{eq:s-wave-rate}
 \end{align} 
where  $m_{\rm DM}, n_{\rm DM}=n_{\rm DM,0}(1+z)^3$ denote the DM mass and redshift-dependent number density, respectively, with $n_{\rm DM,0}$ the present-day value and $\Omega_{\rm DM}h^2\approx0.12$~\cite{Planck:2018vyg}.  From Eq.~\eqref{eq:f-param} to Eq.~\eqref{eq:s-wave-rate}, we have assumed a redshift-independent, $s$-wave annihilation cross section, with $6\times 10^{-26}\text{cm}^3/\text{s}$ the value expected from   the standard weekly-interacting-massive-particle (WIMP)    paradigm~\cite{Kolb:1990vq}.
The parameter $f_{\rm EM}$ is defined to characterize the energy deposition into electromagnetic cascade, or the ratio of electromagnetic deposition to the total energy release at a given redshift~\cite{Padmanabhan:2005es}. 

The computation of $f_{\rm EM}$ is, by definition, DM model dependent, so  the bounds on $\langle\sigma v\rangle$ depend  on the details of $f_{\rm EM}$. In earlier studies, $f_{\rm EM}$ was simply treated as a constant, which corresponds to the on-the-spot approximation  and was applied to investigate the impact of DM annihilation at recombination~\cite{Padmanabhan:2005es}.  
 The redshift-dependent  behavior of  $f_{\rm EM}$, partly due to the photon \textit{transparency window}, was later considered in  Ref.~\cite{Slatyer:2009yq}, though the variation of $f_{\rm EM}(z)$ is   within $\mathcal{O}(10\%)$ for $10\lesssim z\lesssim 1100$.  For  $z\gg z_{\rm rec}$,  $f_{\rm EM}$ approaches the redshift-independent limit, since the   energy-transfer
rates for electrons and photons  become much rapid with respect  to the Hubble expansion. 
It should be mentioned that  $f_{\rm EM}$ was originally defined to investigate the impact of   electromagnetic injection (electrons and photons) at recombination and reionization. It can be  applied not only to direct electron/photon injection, but also to secondary production from heavier SM particles and even short-lived hidden particles.\footnote{Theses hidden particles have lifetimes too short to deposit significant energy from DM annihilation so that prompt decay and radiation to electrons and photons  occur. On the other hand, in some DM scenarios with scalar mediators, the annihilation channel to electrons can be suppressed by the  electron mass, and the direct electron injection rate could be much smaller than secondary production. }

For primordial CMB distortions formed at $z\gg z_{\rm rec}$, $f_{\rm EM}$  can still be approximately applied to investigate the impact on $\mu$  and $y$ distortions in a model-independent manner.  
 While  $f_{\rm EM}$  also includes the situation of  electron-positron injection,   most of direct  $e^\pm$ production or  secondary cascade from prompt decay will lose their energy to photons by   inverse Compton scattering (strong Compton cooling) on background photons,  and the boosted photons then redistribute their energy via Compton scattering, leading to the $\mu$  and $y$ distortions.  From now on, we will not distinguish the difference between $e^\pm$ and photon injection. The discussion of energy injection to $e^\pm$ pairs and its difference from photon injection is presented in  Appendix~\ref{app:e-channel}.

The nearly common dependence of recombination, reionization and primordial spectrum distortions on $f_{\rm EM}\langle \sigma v\rangle$  leads to the fate that the   constraints from CMB observations severely limit the sizes of  primordial $\mu$  and $y$ distortions that can be created by DM $s$-wave annihilation. To see this, we integrate Eq.~\eqref{eq:s-wave-rate} with Eq.~\eqref{eq:visibility-fun}, giving rise to  
\begin{align}\label{eq:s-wave-mudis}
	\mu&\approx 1.4\int_{z_{\mu}}^{z_{\mu y}}\frac{1}{\rho_\gamma}\frac{\delta \rho}{\delta t}\frac{dt}{dz}dz\approx 8.3\times 10^{-10}\left(\frac{f_{\rm EM}\langle\sigma v\rangle/m_{\rm DM}}{6\times 10^{-28}~\text{cm}^3/\text{s}/\text{GeV}}\right),
	\\[0.3cm]
		y&\approx 0.25 \int_{z_{\mu y}}^{z_{\rm rec}}\frac{1}{\rho_\gamma}\frac{\delta \rho}{\delta t}\frac{dt}{dz} dz\approx 0.18 \mu\,,\label{eq:s-wave-ydis}
\end{align}
with the background photon energy density   given by
\begin{align}
\rho_\gamma\approx 0.26\times 10^{12}\left(\frac{1+z}{1000}\right)^4~\text{eV}/\text{cm}^3\,.
\end{align} 
 
The  above  discussion assumes   a redshift-independent cross section $\langle\sigma v\rangle$,  but   for a broad class of DM scenarios, $\langle\sigma v\rangle$ can depend on the redshift and hence lead to   different  total energy injection. One typical example is the Sommerfeld enhancement~\cite{Sommerfeld:1931qaf}. In the simple case,  DM $s$-wave annihilation is modified by
\begin{align}
\langle\sigma v\rangle=S \langle\sigma v\rangle_0\,,
\end{align}
with the Sommerfeld-enhanced factor $S$ depending on the    relative velocity of annihilating DM,  and $\langle\sigma v\rangle_0$ corresponding to the cross section in the Born approximation without the Sommerfeld effect. The redshift dependence then is determined by the relative velocity and the history of DM kinetic decoupling.  For  $S\propto 1/v$, the impacts at recombination and reionization become larger such that   smaller primordial distortions are allowed~\cite{Zavala:2009mi}. 
In fact, significant Sommerfeld effects are not so easy to realize. The effect would  hint at a long-range force (light hidden force mediator) and a large coupling between DM and the mediator~\cite{Arkani-Hamed:2008hhe,Feng:2009hw,Feng:2010zp}. In this case, the dominant DM annihilation channel is the hidden mediator. If the hidden mediator is long-lived, the bound  from  CMB observations and the primordial distortions  cannot be directly correlated, since  the CMB bound  extracted corresponds to  late-time energy injection from hidden particle decay, which is  not equal to   early-time  energy injection from  DM annihilation.  

The Sommerfeld effect could also be velocity-dependent suppression. In this case, later energy injection would be smaller than in earlier injection such that the constraints from CMB observations can be significantly relaxed.   We leave this   possibility for future studies,  and in the following turn to a more typical scenario of redshift-dependent  DM annihilation: $p$-wave dominated annihilation.

\subsection{Annihilation via $p$-wave}\label{sec:mudist-pwave}
For DM $p$-wave annihilation to electromagnetic cascade, the thermally averaged cross section  can be written as
\begin{align}\label{eq:p-wave-cs}
f_{\rm EM}\langle \sigma v\rangle \equiv b \langle v^2\rangle=b\frac{\int d^3 p_1 d^3 p_2 f_1 f_2 v^2}{\int d^3 p_1 d^3 p_2 f_1 f_2}=6  \frac{T_{\rm DM}}{m_{\rm DM}}b\,,
\end{align} 
where $b$ denotes the redshift-independent constant and $v^2=|\vec{p}_1/E_1-\vec{p}_2/E_2|^2\approx (p_1^2+p_2^2-2\vec{p}_1\cdot \vec{p}_2)/m_{\rm DM}^2$ is the relative squared velocity of the annihilating DM pair. $T_{\rm DM}$ is conventionally defined as the  effective DM \textit{temperature} (see e.g. Ref.~\cite{Bringmann:2006mu}):
\begin{align}\label{eq:DM-temp}
	T_{\rm DM}\equiv \frac{1}{3m_{\rm DM}n_{\rm DM} }\int \frac{d^3 p}{(2\pi)^3}p^2 f_{\rm DM}(p)\,,
\end{align}
where $f_{\rm DM}(p)$ is an arbitrary (isotropic) DM distribution function and the above definition holds trivially for the Maxwell-Boltzmann distribution. Unlike $s$-wave annihilation, to study the cosmological impacts of $p$-wave annihilation, such as BBN, CMB, structure formation, and the primordial CMB spectrum distortions, one should know the effective DM temperature, or equivalently the DM distribution function, at different redshifts.  For nonthermal DM,   determination of the   distribution function is usually complicated and time consuming, and   depends on the details of DM-SM interactions. For such nonthermal DM, therefore,   determination of primordial CMB spectrum distortions should be done  in a case-by-case manner. 

\begin{figure}
	\centering
	
	\includegraphics[scale=0.7]{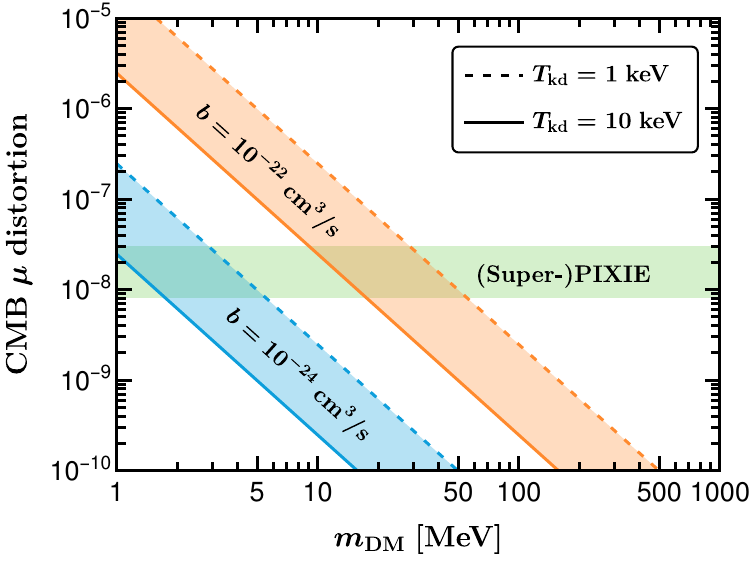} 
	
	\caption{\label{fig:CMBmu} Predictions of the $\mu$-distortion from $p$-wave DM annihilation to $e^\pm$/$\gamma$, with a kinetic decoupling temperature ranging from 1~keV to 10~keV. Two   annihilation  cross sections are chosen for references, corresponding to the order predicted by  thermal WIMP DM ($b=10^{-24}\,\text{cm}^3/\text{s}$) and allowed by current observations ($b=10^{-22}\,\text{cm}^3/\text{s}$), respectively. The observational sensitivities from (Super-)PIXIE  missions $(0.8-3)\times 10^{-8}$ are shown in the horizontal band. }
\end{figure}

 To be as model-independent as possible, let us turn to a simpler class of DM models. Suppose DM keeps  in   kinetic equilibrium with the background plasma in   earlier times. Then before kinetic decoupling, we have $T=T_{\rm DM}$. After kinetic decoupling at $T_{\rm kd}$, the DM momentum   redshifts  with the scale factor $a(t)$  as $p_{\rm DM}\propto 1/a$.  Given $n_{\rm DM}\propto 1/a^{3}$, we can read from Eq.~\eqref{eq:DM-temp}  that $T_{\rm DM}\propto 1/a^{2}$. Therefore, we can write
\begin{align}\label{eq:v2}
	 \langle v^2\rangle&= \frac{6T}{m_{\rm DM}}\theta(T-T_{\rm kd})+ \frac{6T^2}{m_{\rm DM}T_{\rm kd}}\theta(T_{\rm kd}-T)
	 \\[0.3cm]
	 &=\frac{6T_0}{m_{\rm DM}}\left[(1+z)\theta(z-z_{\rm kd})+ \frac{(1+z)^2}{1+z_{\rm kd}}\theta(z_{\rm kd}-z)\right]
\end{align}
in the approximation of  instantaneous kinetic decoupling at $T_{\rm kd}\equiv T_0 (1+z_{\rm kd})$. 
 Note that for non-instantaneous kinetic decoupling,  the DM temperature should have a smooth transition from the $1/a$ scaling to the $1/a^2$ scaling. It implies that at later times, the DM temperature from non-instantaneous kinetic decoupling is somewhat higher than from instantaneous decoupling, and hence Eq.~\eqref{eq:p-wave-cs} would predict a larger amount of energy injection for a given constant $b$. In this case, the   DM   temperature in the end of kinetic decoupling is different from the photon temperature, so the  parameter $T_{\rm kd}$ used in the following  should be interpreted as the ratio $T_{\rm DM, kd}/T_{\gamma, \rm kd}^2$, with $T_{\gamma, \rm kd}$ the photon temperature in the end of  non-instantaneous kinetic decoupling.

 The energy-transfer rate  from $p$-wave annihilation is given by Eq.~\eqref{eq:f-param} with the $p$-wave cross section given by Eq.~\eqref{eq:p-wave-cs}. Similar to Eqs.~\eqref{eq:s-wave-mudis}-\eqref{eq:s-wave-ydis}, 
we can derive the distortions from  $p$-wave annihilation as
\begin{align}\label{eq:p-wave-mudis}
 \mu&\approx 5.3\times 10^{-8} \,\mathcal{G}_1(z_i)\left(\frac{b}{  10^{-21}~\text{cm}^3/\text{s}}\right)\left(\frac{100~\text{MeV}}{m_{\rm DM}}\right)^2,
	\\[0.2cm]
	y&\approx  9.4\times 10^{-9} \,\mathcal{G}_2(z_i)\,\left(\frac{b}{  10^{-21}~\text{cm}^3/\text{s}}\right)\left(\frac{100~\text{MeV}}{m_{\rm DM}}\right)^2\,,\label{eq:p-wave-ydis}
\end{align}
 where $\mathcal{G}_{1,2}(z_i)$ encode the relative redshift among $z_{\mu}, z_{\mu y}, z_{\rm kd}$. In the approximation of  $z_\mu\gg z_{\mu y}$ and $z_{\rm kd}, z_{\mu y}\gg z_{\rm dec}$, we have  
 \begin{align}
 \mathcal{G}_1(z_i)&\approx	\left(\frac{z_{\rm kd}}{2\times 10^6}\right)\left[\left(\frac{z_\mu^2}{z_{\rm kd}^2}-1\right)\theta_1+2\left(\frac{z_\mu}{z_{\rm kd}}-1\right)\theta_2+1-\frac{z_{\mu y}^2}{z_{\rm kd}^2}\right],
 \\[0.2cm]
 \mathcal{G}_2(z_i)&\approx\left(\frac{z_{\rm kd}}{2\times 10^6}\right) \left(\frac{z_{\mu y}}{z_{\rm kd}}\right)^2,
 \end{align}
with $\theta_1\equiv \theta(z_{\rm kd}-z_\mu), \theta_2\equiv \theta(z_{\mu}-z_{\rm kd})$. 
 For $z_{\rm kd}\gg z_\mu$, $\mathcal{G}_1\approx 0$, and for $z_{\mu y}\ll z_{\rm kd}\ll z_\mu$, $\mathcal{G}_1\approx 2$. It implies  that significant $\mu$  and $y$ distortions can only be generated with a kinetic decoupling temperature not far above 0.5~keV. 
 For $z_{\rm kd}>z_{\mu y}$,  corresponding to a kinetic decoupling temperature above 10~eV, one can check that     a $\mu$-distortion reaching   $5.3\times 10^{-8}$ would imply the maximal $y$-distortion:  $1.2\times 10^{-10}$,  too small to be observable.  
 
 In Fig.~\ref{fig:CMBmu}, we show the predictions of $\mu$-distortions from DM $p$-wave annihilation, with two representative kinetic decoupling temperatures. We take two annihilation cross sections for comparison, one from the standard WIMP DM, $b=10^{-24}\,\text{cm}^3/\text{s}$, and the other  reaching the order of  current upper bounds, $b=10^{-22}\,\text{cm}^3/\text{s}$.  It can be inferred that observable $\mu$-distortions favor a DM mass below 50~MeV.
 
It should be mentioned again that the approximation of using  Eq.~\eqref{eq:visibility-fun}   will not include the intermediate and nonthermal relativistic     distortions  formed around $z\sim z_{\mu y}$. However,  since the dominant contribution from $p$-annihilation to the
$\mu$-distortion resides at higher redshift $z> z_{\mu y}$, it clarifies the purpose of focusing on  the most profound $\mu$-distortion induced by DM $p$-wave annihilation.

From the redshift dependence of $p$-wave annihilation,   the impacts on the late-time recombination and reionization are much suppressed with respect to the primordial distortions formed at $z\gg z_{\rm rec}$, since $T^2\ll m_{\rm DM} T_{\rm kd}$.
However, for some   DM masses and kinetic decoupling temperatures, the DM velocity during the   $\mu$-era can be comparable with that during/after the BBN epoch and with the local DM velocities at typical  galaxies: $v_{\rm DM}\sim 10-100~\text{km}/\text{s}$. Therefore, one can infer that when a significant $\mu$-distortion is created by DM $p$-wave annihilation,   bounds from BBN and astrophysical observations are also at play.

 \section{Bounds on distortion detection regimes}\label{sec:detection}
 \subsection{Limitation of $s$-wave annihilation}
  For $s$-wave annihilation,  the   current bound from Planck TT, TE, EE+lowE+lensing+BAO reads~\cite{Planck:2018vyg}
 \begin{align}\label{eq:Planck-bound}
 	f_{\rm EM}\frac{\langle\sigma v\rangle}{m_{\rm DM}}<3.2\times 10^{-28}\,\text{cm}^3\cdot \text{s}^{-1}\cdot \text{GeV}^{-1}\,.
 \end{align}
From Eqs.~\eqref{eq:s-wave-mudis}-\eqref{eq:s-wave-ydis}, it  indicates that the primordial $\mu$  and $y$ distortions from DM $s$-wave annihilation can only reach 
 \begin{align}
 	\text{Planck~limit}:~\quad \mu\simeq 4.4\times 10^{-10}\,, \quad y\simeq 8.1\times 10^{-11}\,.
 \end{align} 
 These distortions are not expected to differ dramatically when refined approximations beyond Eq.~\eqref{eq:visibility-fun} are used. Therefore, we conclude that primordial distortions from DM $s$-wave annihilation are below the (Super-)PIXIE sensitivity conservatively by one to two orders of magnitude. 
 Improvement of the bound given in~\eqref{eq:Planck-bound} from future CMB measurements is not expected to increase significantly and is nearly saturated by current observations~\cite{Green:2018pmd}. Indeed, as   anticipated explicitly in Ref.~\cite{Cang:2020exa},  the forecast exclusion limit for annihilation to electrons and photons  is at $\langle\sigma v\rangle/m_{\rm DM}\sim 10^{-29}\,\text{cm}^3/\text{s}/\text{GeV}$. The largest $\mu$  and $y$ distortions allowed by this CMB forecast exclusion limit are then given by
 \begin{align}
 	\text{future~CMB~limit}:~\quad \mu\simeq 1.4\times 10^{-11}\,, \quad y\simeq 2.5\times 10^{-12}\,.
 \end{align} 
 Even if   future sensitivities  can be enhanced to reach the above level,  it would be quite challenging to distinguish these primordial distortions generated by  DM $s$-wave annihilation from sources within the $\Lambda$CDM model. In particular, it could readily  be overwhelmed respectively by the $\mu$-distortion from  the damping of primordial small-scale perturbations\footnote{If the spectrum distortion from Silk damping is not   scale invariant at multipoles higher than what is probed by Planck, the amplitude might fall off and thus induce smaller spectrum distortions~\cite{Cyr:2023pgw}.}
 and   the $y$-distortion generated during the epochs of reionization and structure formation.
 
 It is worth emphasizing  that Eq.~\eqref{eq:f-param} does not assume a special thermal history for DM, the DM distribution function, or  the detailed mechanism for freezing the DM   abundance in   the early epoch. Instead, it takes the minimal assumptions: DM is absolutely stable and   DM pair annihilation to electromagnetic species occurs in the nonrelativistic regime. Therefore, the above conclusions hold for a broad class of simple DM scenarios with $s$-wave dominated annihilation to electromagnetic particles, which is  particularly applicable to the standard WIMP DM, freeze-in DM~\cite{McDonald:2001vt,Kusenko:2006rh,Petraki:2007gq,Hall:2009bx,Bernal:2017kxu}, and some secluded WIMP scenarios~\cite{Pospelov:2007mp,Evans:2017kti,Kanemura:2023jiw} if the hidden mediators have prompt decay to charged particles.
 
 \subsection{Limitation of $p$-wave annihilation}
 \textbf{BBN}.---When DM keeps in thermal equilibrium with the SM plasma down to MeV-scale temperatures, observations of the   primordial light elements formed at the BBN epoch and the effective neutrino number $N_{\rm eff}$ would set strong bounds on  the DM mass below 10~MeV~\cite{Serpico:2004nm,Ho:2012ug,Boehm:2013jpa,Nollett:2013pwa,Nollett:2014lwa,Kawasaki:2015yya,Depta:2019lbe,Sabti:2019mhn,Sabti:2021reh}.\footnote{See, however, Ref.~\cite{Berlin:2017ftj} for an exception.} On the one hand,  light thermal DM itself can contribute   to the energy budget of the early universe  and  hence accelerate the expansion at MeV-scale temperatures. On the other hand, the DM annihilation product, either being electron-positron pairs, photons, or neutrinos, can affect the synthesis of light elements and neutrino decoupling.  The effect from annihilation products  could be  stronger than that from  light DM itself, depending on the ratio of electromagnetic annihilation to neutrino injection. Since significant CMB spectrum distortions favor larger electromagnetic annihilation than neutrino injection, we focus on the situation where DM annihilation to neutrinos is subdominant. In this case, 
  we apply the following mass bound obtained in Refs.~\cite{Depta:2019lbe,Sabti:2019mhn,Sabti:2021reh} with  fermionic DM dominantly  annihilating to electron-positron pairs or photons,\footnote{Note that $p$-wave annihilation is in general expected from fermionic DM.}
   \begin{align}
 	\text{BBN+Planck~bound}:\quad m_{\rm DM}>7~\rm MeV\,.
 \end{align}
 
 Given the above bound, an important observation  from Eq.~\eqref{eq:p-wave-mudis} arises. For an observable $\mu$-distortion reaching $1\times 10^{-8}$, we have  $b\gtrsim 4.6\times 10^{-25}~\text{cm}^3/\text{s}$. This lower limit is  comparable  with   the annihilation cross section predicted by the standard   WIMP DM freeze-out. To see this, recall that  the  thermally averaged cross section for WIMP DM with  a freeze-out temperature $T_{\rm fo}\approx m_{\rm DM}/25$ is given by
 \begin{align}
 \langle \sigma v\rangle_{\rm WIMP}\approx 6.4\times 10^{-26}~\text{cm}^3/\text{s}\,.
 \end{align}
For $p$-wave annihilation, the DM relative velocity during freeze-out is   at $\langle v^2\rangle= 6T_{\rm fo}/m_{\rm DM}\approx 0.24$, indicating a thermal freeze-out annihilation cross section $b\simeq 5.3\times10^{-25}~\text{cm}^3/\text{s}$. Therefore,    observable $\mu$ distortions  generated by  thermal DM $p$-wave annihilation following   the standard freeze-out paradigm can reach at most around $1\times 10^{-8}$. Alternative mechanisms for freezing the DM abundance could generate a larger $\mu$ distortion, but
 we will not specify the   details of the  model-building aspects. Instead,  we   focus on    residual DM annihilation, assuming that the DM abundance is frozen, then redshifts and finally  leads to the present-day relic density.\footnote{As can be inferred from  Eq.~\eqref{eq:s-wave-rate},  it can also be the case that  the annihilation cross section as large as $b\simeq 4.6\times 10^{-25}~\text{cm}^3/\text{s}$ makes the final DM relic density too small to account for the observed value, so that the DM species is only a small component of the whole DM abundance. However, it brings about an  arbitrary degree of freedom and we will not consider this situation further.}

 \textbf{Photodisintegration}.---The BBN constraints discussed above become less severe when DM is heavier than 10~MeV. Nevertheless,     electromagnetic energy injection  after the end of the BBN  can readily reach the photodisintegration threshold such that   high-energy photons can disintegrate the light elements synthesized during the BBN epoch~\cite{Reno:1987qw,Depta:2019lbe}.       In this mass range, the   photodisintegration  effect  depends on whether photons are produced directly or via secondary cascade from DM annihilation. 
 For a   qualitative grasp, let us  consider the impact on deuterium disintegration, $\text{D}+\gamma\to p+n$, with energy threshold $E_{\rm D}\approx 2.22$~MeV. The direct photon production    from DM annihilation, $\text{DM}+\text{DM}\to \gamma+\gamma$,  generates high-energy photons $E_\gamma\sim m_{\rm DM}$. After  the photon spectrum is formed via Compton scattering,   a significant portion of the  photon abundance exists at $E_\gamma>E_{\rm D}$ such that the disintegration reaction of the fragile deuterium occurs frequently. However, when the boosted photon energy are inherited from secondary cascade, e.g.,  $\text{DM}+\text{DM}\to e^++e^-$ followed by   inverse Compton scattering $e+\gamma_{\rm bg}\to e+\gamma$ on background photons $\gamma_{\rm bg}$, the produced photons are softer, $E_\gamma< m_{\rm DM}$. In this case, the deuterium disintegration  rate becomes smaller, since the photon distribution in the high-energy band becomes suppressed.  Therefore, the photodisintegration   effect depends on the high-energy band of   photons and hence   on whether hard or soft photons are injected. 
 
 The BBN photodisintegration from $p$-wave annihilation also depends on the relative velocity of the annihilating DM pair. Therefore, the constraints due to the  photodisintegration effects would  depend on the DM kinetic decoupling temperature. As can be inferred from Eq.~\eqref{eq:p-wave-mudis}, a lower $T_{\rm kd}$ is favored to   generate a larger $\mu$-distortion, but it also leads to stronger constraints from   measurements of light elements. This is because  a lower $T_{\rm kd}$ indicates a larger $\langle v^2\rangle$ and hence a larger energy injection rate at the photodisintegration epoch ($T\lesssim 10$~keV).

\textbf{Kinetic decoupling}.---The $\mu$-distortion estimated in Sec.~\ref{sec:mudist-pwave} depends on the kinetic  decoupling temperature.  For non-standard WIMP DM,    kinetic decoupling could be much later than  chemical decoupling.  Nevertheless, a late kinetic decoupling temperature for DM can suppress the structure formation at small   scales and induce a  cutoff in the primordial power
spectrum of density perturbations~\cite{Green:2005fa}.
Under the definition of the DM temperature given by Eq.~\eqref{eq:DM-temp}, the small-scale cutoff in the DM halo mass function  reads~\cite{Vogelsberger:2015gpr}
\begin{align}\label{eq:Mcut-kd}
	M_{\rm cut, kd}=5\times 10^8\left(\frac{1\, \rm keV}{T_{\rm kd}}\right)^{3}h^{-1}M_\odot\,,
\end{align}
 with $h=(H_0/100) \cdot\text{km}^{-1}\cdot \text{s}\cdot\text{Mpc}\approx 0.67$ the dimensionless Hubble constant~\cite{Planck:2018vyg} and $M_\odot$ the solar mass.
 
A low $T_{\rm kd}$ can be realized by some couplings between DM and relativistic particles (photons, neutrinos, or some light dark species),  which was considered previously in Refs.~\cite{vandenAarssen:2012vpm,Chu:2014lja,Bringmann:2016ilk}  as an alternative possibility to   alleviate the \textit{small-scale crisis}~\cite{Tulin:2017ara,Bullock:2017xww}.  
	However, a lower kinetic decoupling temperature hints at a relatively large DM-radiation interaction and could also lead to non-negligible DM self-interactions, where  constraints from observations of  galaxy and galaxy clusters should be further taken into account. In particular, observations from Milky Way substructures disfavor a kinetic decoupling temperature $T_{\rm kd}<0.7$~keV~\cite{Benito:2020avv} if  DM couples to a dark radiation bath. 
	
	 On the other hand,   it is known that warm DM scenarios can also  suppress the structure formation at small   scales, with the cutoff function   fitted by~\cite{Vogelsberger:2015gpr} 
	 \begin{align}\label{eq:Mcut-WDM}
	 	M_{\rm cut, WDM}= 10^{11}\left(\frac{1\, \rm keV}{ m_{\rm WDM}}\right)^{4}h^{-1}M_\odot\,.
	 \end{align}
	Some lower mass bounds of warm DM have been derived by observations from      Lyman-$\alpha$ data~\cite{Irsic:2017ixq}  ($m_{\rm WDM}>5.3$~keV)  and  from   Milky Way satellite galaxies~\cite{DES:2020fxi} ($m_{\rm WDM}>6.5$~keV).
	Using these mass bounds, we can infer a lower limit of the kinetic decoupling temperature by requiring that  $M_{\rm cut, kd}<M_{\rm cut, WDM}$, which
	gives the following conservative limit:
\begin{align}\label{eq:Tkd_bound}
T_{\rm kd}\gtrsim 1~\rm keV\,.
\end{align} 

When photon energy injection from DM annihilation  is concerned, it is assuring to  check whether the allowed elastic   DM-photon scattering cross section  enables  a kinetic decoupling temperature to satisfy  condition~\eqref{eq:Tkd_bound}. To this end,  we   apply the  constraint of a constant DM-photon scattering cross section from the CMB angular power spectrum~\cite{Wilkinson:2013kia},
\begin{align}\label{eq:sigma_gammaDM}
	\sigma_{\gamma \text{DM}}<8\times 10^{-32}\left(\frac{m_{\rm DM}}{100\,\text{MeV}}\right) \text{cm}^2
\end{align}
to determine the kinetic decoupling temperature via
\begin{align}
n_{\gamma}\langle\sigma v\rangle_{\gamma \text{DM}}=H(T_{\rm kd})\,,
\end{align}
with $v\approx 1$ the relative velocity between the incoming photon and DM,  and $n_\gamma\approx 0.24 T^3$ the photon number density.  One can check that   condition~\eqref{eq:Tkd_bound} is consistent with~\eqref{eq:sigma_gammaDM}.

\textbf{Cosmic electron-positron rays}.--- 
When primordial distortions are created by  DM annihilation to electron-positron pairs, the continuing annihilation will contribute to cosmic electron-positron rays. Given Eqs.~\eqref{eq:v2} and~\eqref{eq:Tkd_bound},   the DM velocity  during the formation of  primordial  distortions   is  $\langle v^2\rangle\lesssim 10^{-3}/(m_{\rm DM}/\rm MeV)$, which, for $m_{\rm DM}\gg 1$~MeV, is comparable  with the local DM velocity in Milky-Way-size galaxies: $v_{\rm DM}\simeq 220~\text{km}/\text{s}$.  It implies that a larger $b$ for primordial CMB spectrum distortions will also lead to abundant $e^\pm$ rays that shall be observed today.
For DM $p$-wave annihilation, the current bound from cosmic $e^\pm$ rays was derived in Ref.~\cite{Boudaud:2018oya} using two propagation models for the cosmic rays. The strong   re-acceleration model is no longer supported by current observation data~\cite{Genolini:2021doh} while the negligible re-acceleration  model can  still be used as a relevant proxy to estimate the constraints on DM $p$-wave annihilation from e.g. Voyager~\cite{Genolini:2019ewc}. It is worthwhile to mention that, besides the propagation of cosmic rays,  the constraints also depend on the local DM density and velocity distributions, with    uncertainties from modeling the local  DM profiles. To be conservative, we apply the constraints from Ref.~\cite{Boudaud:2018oya} with the negligible re-acceleration  model and the Navarro-Frenk-White DM density profile.

\textbf{Diffuse photon background}.---Besides the cosmic $e^\pm$ rays, DM annihilation to photons in the (extra-)galaxies   contributes to the   
diffuse gamma-ray background. The photons from DM annihilation can either be generated via direct production or by final-state radiation of $\text{DM}+\text{DM}\to e^++e^-$. For the former, we apply the bounds from Ref.~\cite{Boddy:2015efa} for $m_{\rm DM}\in [1,200]$~MeV and from Ref.~\cite{Albert:2014hwa} for $m_{\rm DM}>200$~MeV, while for the latter, we adopt the reuslts from Ref.~\cite{Essig:2013goa}.   In either case, the bounds extracted are given in terms of $\langle \sigma v\rangle$, so the  corresponding limits on $b$, as relevant to CMB spectrum distortions, would depend on the local DM velocity.  For typical galaxies, we take $v_{\rm DM}\simeq 220~\text{km}/\text{s}$,  as also adopted in Ref.~\cite{Essig:2013goa}.


 \begin{figure}
 	\centering
 	
 	\includegraphics[scale=0.525]{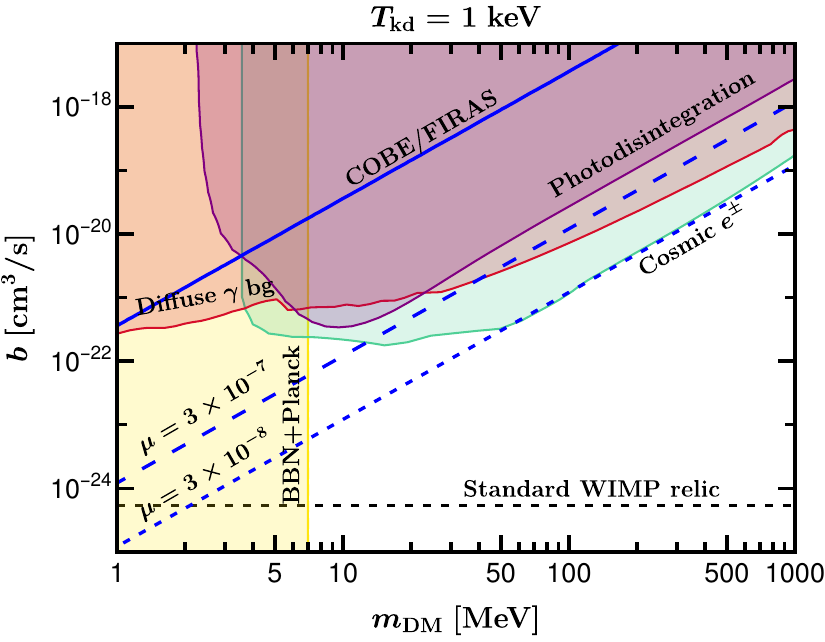}\quad
 	\includegraphics[scale=0.525]{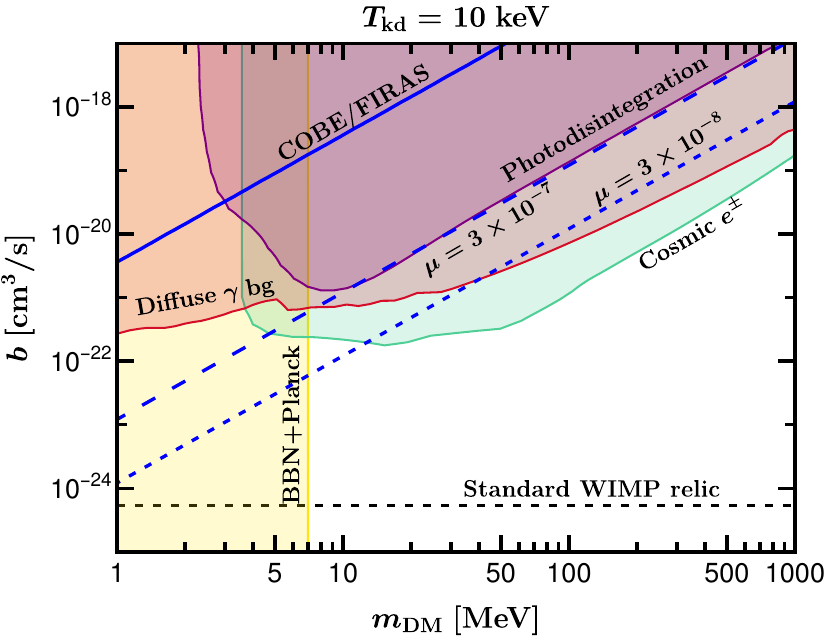} 
 	
 	\caption{\label{fig:bound-ee} Predictions of the CMB $\mu$-distortion under the  current bounds of $p$-wave DM annihilation to $e^\pm$ pairs from photodisintegration~\cite{Depta:2019lbe}, cosmic $e^\pm$ rays~\cite{Boudaud:2018oya} and diffuse photon background (Diffuse $\gamma$ bg)~\cite{Essig:2013goa}. The lower mass bound $m_{\rm DM}>7$~MeV is set by BBN+Planck~\cite{Depta:2019lbe,Sabti:2019mhn,Sabti:2021reh}. Two kinetic decoupling temperatures are chosen, with the lower value (1~keV) inferred from the Lyman-$\alpha$~\cite{Irsic:2017ixq} and structure formation~\cite{DES:2020fxi}. The current bound of the CMB $\mu$-distortion from COBE/FIRAS is also shown. The thermal annihilation cross section  from standard WIMP DM, $b\approx 5.3\times 10^{-25}~\text{cm}^3/\text{s}$,  is shown by the horizontal line. }
 \end{figure}

\begin{figure}
	\centering
	
	\includegraphics[scale=0.526]{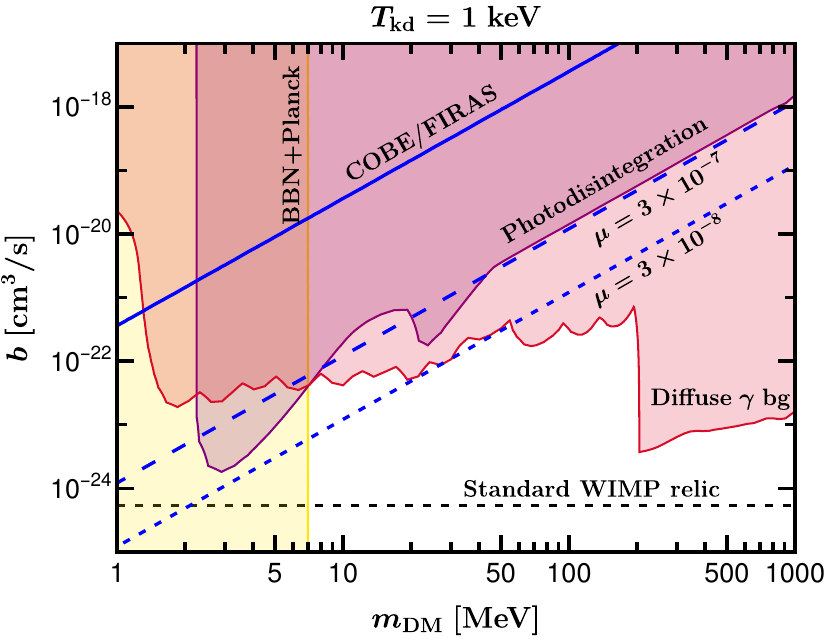}\quad
	\includegraphics[scale=0.526]{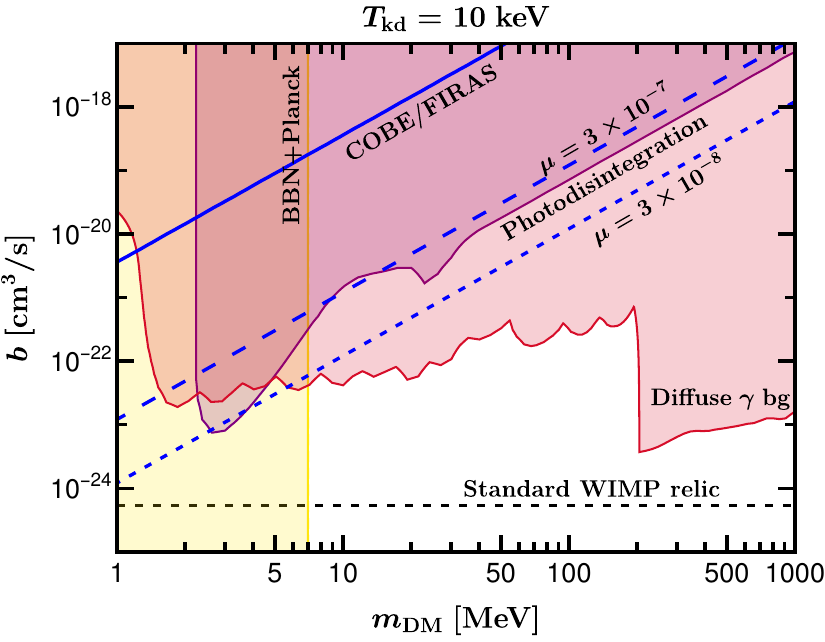} 
	\caption{\label{fig:bound-gg}Predictions of the CMB $\mu$-distortion under the  current bounds of $p$-wave DM annihilation to photons  from photodisintegration~\cite{Depta:2019lbe}, and diffuse photon background (Diffuse $\gamma$ bg)~\cite{Boddy:2015efa,Albert:2014hwa}. The lower mass bound $m_{\rm DM}>7$~MeV is set by BBN+Planck~\cite{Depta:2019lbe,Sabti:2019mhn,Sabti:2021reh}. Two kinetic decoupling temperatures are chosen, with the lower bound (1~keV) inferred from the Lyman-$\alpha$~\cite{Irsic:2017ixq} and structure formation~\cite{DES:2020fxi}. The thermal annihilation cross section  from standard WIMP DM, $b\approx 5.3\times 10^{-25}~\text{cm}^3/\text{s}$,  is shown by the horizontal line.}
\end{figure}

We show in Figs.~\ref{fig:bound-ee} and~\ref{fig:bound-gg} the current  constraints of  DM $p$-wave annihilation to $e^+e^-$ and photons. Besides, we also show   the current bound of the CMB $\mu$-distortion from the COBE/FIRAS measurement  and   two reference $\mu$-distortions predicted by   DM annihilation.  It can be read from the figures that the current COBE/FIRAS  constraint is not competitive with other bounds set by BBN and astrophysical observations.
Note that for the $\mu$-distortion,  we do not distinguish the   injection sources from $e^+e^-$ and photon channels, since     $e^\pm$  injection would lose most of   energy to photons via repeated inverse Compton scattering. See Appendix~\ref{app:e-channel} for  a more detailed discussion.

For energy injection to the $e^\pm$ channel, the most severe bound comes from cosmic $e^\pm$ rays for $m_{\rm DM}>7$~MeV. In general, we can see that for a $\mu$-distortion larger than the forecast (Super-)PIXIE sensitivities: $\mu=3\times 10^{-8}$, the constraint from cosmic $e^\pm$ rays   permits a DM mass  below $50$~MeV for  $T_{\rm kd}>1$~keV.  In particular, a $\mu$-distortion as large as $3\times 10^{-7}$   is feasible for $m_{\rm DM}\simeq 10$~MeV and $T_{\rm kd}=1$~keV.  As the kinetic decoupling temperature increases to 10~keV, only a DM mass around 10~MeV is allowed to generate $\mu\gtrsim 3\times 10^{-8}$.

For  energy injection to the photon channel, the most severe bound comes from diffuse photon background for $m_{\rm DM}>7$~MeV. Similar to the  $e^\pm$ channel,  a $\mu$-distortion larger than the forecast (Super-)PIXIE sensitivities is   permitted for a DM mass below 50~MeV. However, a $\mu$-distortion  larger than $3\times 10^{-7}$ is basically disfavored by BBN, photodisintegration and diffuse photon background. For $3\times 10^{-8}\lesssim \mu\lesssim 3\times 10^{-7}$, the inferred DM mass is restricted  in a narrow range: 7--50~MeV, and the kinetic decoupling temperature should be as low as 1~keV.

In both cases, the favored DM mass resides at the order of 10~MeV and the required kinetic decoupling temperature should be around 1~keV. It 
would be worthwhile to wait for the improved measurements of the helium-4 abundance and baryon density in  the upcoming     CMB  experiments, which will be able to probe the DM mass up to 15~MeV~\cite{Sabti:2019mhn}. On the other hand, the improved constraints from the Lyman-$\alpha$ and structure formations at galactic scales can hopefully probe a $T_{\rm kd}$ up to 9~keV~\cite{Benito:2020avv}, providing complementary tests of the CMB $\mu$-distortion signal from DM $p$-wave annihilation.

 \section{Conclusion}\label{sec:con}
This paper has demonstrated the observability of primordial CMB spectrum distortions from simple DM $s$- and $p$-wave annihilation to electromagnetic species, under the constraints from CMB, synthesis and photodisintegration of primordial light elements during and after the BBN epoch, the impact  of late kinetic decoupling on structure formation, the observation of cosmic electron-positron rays, and the diffuse photon background.  

For $s$-wave annihilation, the severe bounds from current CMB observations disfavor any detectable $\mu$ and $y$ distortions under the observational limits of  the  (Super-)PIXIE, PRISM and Voyage 2050 programs,  while  only $\mu\simeq 10^{-11}$ and $y\simeq  10^{-12}$ are expected within  the future CMB detection limit. Even if more delicate programs  can reach such sensitivities, it would be much difficult to extract these DM-sourced signals from other standard sources in the  $\Lambda$CDM model. 

For $p$-wave annihilation, CMB constraints become much weaker due to the velocity-dependent annihilation cross section. Nevertheless,    observations of primordial light elements, cosmic rays and gamma rays severely limit the $p$-wave annihilation cross section.  Furthermore, the structure formation also constrains a late kinetic decoupling temperature for DM. Combining these limitation effects, we find that only a narrow window for   observable $\mu$-distortions   remains, which favors a DM mass in the range 10--50~MeV and a kinetic decoupling temperature around 1~keV.  

With the improved sensitivities of measurements in the CMB and structure formation, the DM mass above 10~MeV  and the kinetic decoupling temperature above 1~keV can be probed. These upcoming measurements can therefore provide promising   tests of the DM-sourced  $\mu$-distortion, complementary to  the analysis of     primordial distortion degeneracy  from various sources in the standard $\Lambda$CDM model.

\section*{Acknowledgements}
The author would like to  thank Xun-Jie Xu and Ke-Pan Xie for helpful discussions on $p$-wave annihilation, and Bingrong Yu for frequent discussions on Sommerfeld effects. The author would   also like to thank 
Jens Chluba for useful comments on CMB spectrum distortions and Julien Lavalle for helpful feedback on the propagation models of cosmic rays.
This work was supported in part by the National Natural Science Foundation of China under grant No.~12141501 and also by CAS Project for Young Scientists in Basic Research (YSBR-099).

\appendix
\section{Temperature differences}\label{app:Tdif}

When primordial CMB  spectrum distortions are concerned, there are in general three different temperatures, the perturbed photon temperature $T_\gamma$, the electron temperature $T_e$ and the background temperature $T$. The last is an effective parameterization of the universe temperature  in the radiation-dominated epoch (radiation temperature): 
\begin{align}
H(T)\approx 1.66\sqrt{g_*(T)}T^2/M_{\rm P}\approx H_0 \sqrt{\Omega_{\rm r}} (1+z)^2\,,
\end{align} 
with $g_*(T)\approx 3.38$ after electron-positron annihilation,  $H_0\approx 67.4~\text{km}/\text{s}/\text{Mpc}$~\cite{Planck:2018vyg} the present-day Hubble constant and $\Omega_{\rm r}\approx 9.19\times 10^{-5}$ the   radiation density including the decoupled neutrinos. The time scale of the Hubble expansion at $z_{\rm rec}\lesssim z\lesssim z_{\mu}$ is characterized by 
\begin{align}\label{eq:tau_H}
	\tau_{H}\equiv H_0^{-1} \left[\Omega_{\rm m,0}(1+z)^3+\Omega_{\rm r, 0}(1+z)^4+\Omega_\Lambda \right]^{-1/2}\,,
\end{align}
with the present-day matter density $\Omega_{\rm m,0}\approx 0.32$ and dark energy density $\Omega_\Lambda\approx 0.68$~\cite{Planck:2018vyg}.  While at least two massive neutrinos are expected to be nonrelativistic at $T_0$, it does not affect the expansion at $z>z_{\rm rec}$, so we simply take the current radiation density  as $\Omega_{\rm r,0}=\Omega_{\rm r}\approx 9.19\times 10^{-5}$.

 In the presence of  energy injection, the radiation temperature can be used as a reference temperature for the evolution of perturbations, following the simple redshifting $T=T_0(1+z)$.  Then we can parameterize the photon temperature shift as  $\Delta T=T_\gamma-T$. The  current CMB temperature observed corresponds to $T_{\gamma,0}\equiv T_\gamma(z=0)$, which is   different from $T(z=0)$ due to energy injection. The temperature shift could have a contribution to other observables, e.g., $\mu, y$ distortions, after substituting $T=T_\gamma -\Delta T$ into those observables that are formulated by the reference temperature $T$.    However, for small energy injection concerned here, the difference between $T_{\gamma,0}$ and $T_0$ is  smaller than the current   observational uncertainty: $\delta T_0/T_0\simeq 10^{-4}$~\cite{Fixsen_2009}. Therefore, throughout this paper, we simply set $T_\gamma=T=T_0(1+z)$. Nevertheless, the difference between $T$ and $T_e$ should be taken into account properly when deriving the  $\mu$- and $y$-distortions at leading order, even if $|T-T_e|/T\ll 1$ is maintained throughout the formation of primordial distortions. 
For $z_{\mu y}\lesssim z\lesssim z_{\mu}$, Compton scattering is efficient to maintain $T_\gamma=T_e$, while the difference   $T_e-T$ is at   $\mathcal{O}(\mu)$. To see this, we can  use the fact  that Comptonization is faster than the Hubble expansion such that  energy and particle conservation holds before and after   energy injection. Applying 
\begin{align}
\rho_\gamma(T)+\delta \rho(T)=\rho_\gamma(T_e)(1-a_\mu \mu)\,,\quad   n_\gamma(T)+\delta n(T)=n_\gamma(T_e)(1-b_\mu \mu)\, ,
\end{align}
 we are led to  
\begin{align}\label{eq:Te-T}
	\left(\frac{T}{T_e}\right)^4=(1-a_\mu \mu)(1-\delta \rho/\rho_\gamma)\,, \quad 	\left(\frac{T}{T_e}\right)^3=(1-b_\mu \mu)(1-\delta n/n_\gamma)\,.
\end{align}
 Using  Eq.~\eqref{eq:mu-dist}, we arrive at
 \begin{align}
 	T\approx T_e(1-0.64\delta \rho/\rho_\gamma+0.51\delta n/n_\gamma)\,,
 \end{align}
  so $|T_e-T|/T\simeq \mathcal{O}(\mu)$. 
  
 For $z_{\rm rec}\lesssim z\lesssim z_{\mu y}$, we have used the strong Compton cooling limit and neglected the Hubble expansion to estimate the primordial $y$-distortion, which amounts to  assume that the Compton scattering rate is still fast enough compared to the Hubble expansion. While this overly simplified  analytic approach overestimates the $y$-distortion, it serves for us  to see how large  a  $y$-distortion can be generated by DM $p$-wave annihilation.  Under this treatment, we can estimate the temperature difference $T-T_e$. Similar to Eq.~\eqref{eq:Te-T}, we use Eq.~\eqref{eq:y-dist} to obtain 
 \begin{align}
 	T\approx T_e(1-0.33 \delta n/n_\gamma)\,,
 \end{align}
 and hence the temperature difference is at $\mathcal{O}(y)$.
 It should be emphasized again that  this conclusion is valid only in the   simplified limit  of strong Comptonization. A more detailed analysis of the $y$-distortion can be found e.g., in Ref.~\cite{Chluba:2015hma}.

   \section{$\mu$-distortion from  $\text{DM}+\text{DM}\to e^+e^-$ annihilation}\label{app:e-channel}
   
   %
   %
   
   The energy injection source discussed in Sec.~\ref{sec:pri-dist} can  result either from direct photon injection, or from electron-positron injection followed by rapid inverse Compton scattering.   To see this more explicitly,  let us consider the case of $e^\pm$ injection and the resulting $\mu$-distortion.  Neglecting the double Compton scattering and bremsstrahlung, we can write  the coupled Boltzmann equations of photons and electrons (positrons)  as
   \begin{align}\label{eq:g-Comptonization}
   	\frac{\partial f_\gamma}{\partial t}-Hp_\gamma \frac{\partial f_\gamma}{\partial p_\gamma}&\approx \frac{\delta f}{\delta t}\Big|_{e+\gamma_{\rm bg}\to e+\gamma}-\frac{\delta f}{\delta t}\Big|_{\gamma+e_{\rm bg}\to e+\gamma}\,,
   	\\[0.2cm]
   	\frac{\partial f_e}{\partial t}-Hp_e \frac{\partial f_e}{\partial p_e}&\approx  -\frac{\delta f}{\delta t}\Big|_{e+\gamma_{\rm bg}\to e+\gamma}+\frac{\delta f}{\delta t}\Big|_{\gamma+e_{\rm bg}\to e+\gamma}+\frac{\delta f}{\delta t}\Big|_{e-\text{inj}}\,.
   	\label{eq:e-Comptonization}
   \end{align}
   In the right-hand side of Eq.~\eqref{eq:g-Comptonization}, the first term arises from inverse Compton scattering of injected electrons boosting  the background photons $\gamma_{\rm bg}$, and the second term is the inverse process of boosted photons scattering on the background electrons $e_{\rm bg}$. 
   Similarly, in the right-hand side of Eq.~\eqref{eq:e-Comptonization}, the first term corresponds to the loss rate of injected electrons, i.e., the inverse Compton scattering on background photons,    the second term is the gain rate from boosted photons which scatter on and consequently boost the   background electrons, and the third term is the energy injection source. 
   Note that we have   neglected   photon-energy injection from  $e^\pm$ annihilation, since for $m_{\rm DM}\gg m_e$ relevant here the injected $e^\pm$ pairs carry energy from DM annihilation much higher than   $m_e$, ensuring that most of the injected energy is processed via inverse Compton scattering rather than $e^\pm$ annihilation.\footnote{It should be cautious that when $m_{\rm DM}\sim m_e$, a significant portion of injected energy will be stored in the  $e^\pm$ rest mass energy. If the  $e^\pm$ annihilation timescale  is longer than the Hubble time, the CMB distortions created by $\text{DM}+\text{DM}\to e^++e^-$ annihilation would differ significantly from  $\text{DM}+\text{DM}\to \gamma+\gamma$. }

   By combining Eq.~\eqref{eq:g-Comptonization} and Eq.~\eqref{eq:e-Comptonization}, we arrive at
   \begin{align}\label{eq:dfgdt}
   	\frac{\partial f_\gamma}{\partial t}-Hp_\gamma \frac{\partial f_\gamma}{\partial p_\gamma}&\approx \frac{\delta f}{\delta t}\Big|_{e-\text{inj}}-HC_e \left(1-\frac{d\ln T_e}{d\ln T }\right),
   \end{align}
   where $dT/dt\approx -H T$ is used and $C_e\equiv  E_e e^{-E_e/T_e}/T_e$ results from the Boltzmann distribution  in the limit $E_e\gg m_e\gtrsim \mu_e$, with $\mu_e$ the electron chemical potential.\footnote{The electron chemical potential can be estimated by electric neutrality: $n_{e^-}-n_{e^+}=n_b\simeq 10^{-9}n_\gamma$, with $n_b$ the baryon density mostly composed of protons and helium-4 at $z\in[z_{\mu y},z_\mu]$.} The difference  in the  right-hand side of Eq.~\eqref{eq:dfgdt} has a simple interpretation:   energy  injection to electron-positron pairs has a small portion in increasing the electron temperature.  Then substituting Eq.~\eqref{eq:fmu-dis} into Eq.~\eqref{eq:dfgdt}, we obtain
   \begin{align}\label{eq:mu-dist-2}
   	-\frac{e^{x_e}}{(e^{x_e}-1)^2}\frac{\partial \mu}{\partial t}+H\left(C_\gamma+C_e\right)\left(1-\frac{d\ln T_e}{d\ln T }\right)&\approx \frac{\delta f}{\delta t}\Big|_{e-\text{inj}}\,,
   \end{align}
   with $C_\gamma\equiv x_e e^{x_e}/(e^{x_e}-1)^2$.  For relevant $x_e\in [0.5,105.6]$ and $E_e/T_e\gg m_e/T_e>10^3$, the approximation $C_\gamma\gg C_e$ holds.   Then integrating the above equation respectively over $d^3 p_\gamma E_\gamma$ and $d^3 p_\gamma$, we can check that  Eq.~\eqref{eq:mu-dist-2}   reduces to Eq.~\eqref{eq:dmudt}. Therefore, if energy injection mostly ends up with ultrarelativistic $e^\pm$ products at $z\in [z_{\mu y},z_\mu]$, most of the energy from injected $e^\pm$ will be stored in photons via inverse Compton scattering.

 \section{Rates and cooling timescales}\label{app:rates}
 For easy grasp of Comptonization and thermalization timescales during  the formation of photon  spectrum distortions, we collect here the relevant photon-electron collisions and the timescales with respect to the Hubble time given in Eq.~\eqref{eq:tau_H}.

 \textbf{Compton scattering}.---The Compton scattering rate given in Eq.~\eqref{eq:f-Boltzmann} reads~\cite{Novikov2006}
 \begin{align}
 	\frac{\delta f}{\delta t}\Big|_{\rm C}&= \frac{ n_e \sigma_T T_e}{m_e} \frac{1}{x_e^2}\frac{\partial }{\partial x_e}\left[x_e^4\left(\frac{\partial f_\gamma}{\partial x_e}+f_\gamma+f_\gamma^2\right)\right],
 	\label{eq:C}
 \end{align}
 where $n_e$ is the electron number density and $\sigma_T= 8\pi \alpha_{\rm EM}^2/(3m_e^2)$  the Thomson scattering cross section.
 For kinetic equilibrium  with  $f_\gamma=f_\gamma(x_e,0)$, the Compton scattering rate vanishes. The collision term can  be derived by applying the Fokker-Planck expansion~\cite{Sazonov:1999fy}, with the first term in the right-hand side corresponding to the Doppler effect and the second (third) term to the Compton (induced Compton) scattering recoil. 
 
 The Comptonization timescales considered in the literature usually vary by a constant prefactor. Here we take the largest (absolute) Doppler  term for small $x_e$, i.e., $4e^{x_e}x_e/(e^{x_e}-1)^2$, to be compared with the Hubble friction term $-Hx_e \partial f_\gamma/\partial x_e=H e^{x_e}x_e/(e^{x_e}-1)^2$ in the kinetic equilibrium limit. We then arrive at 
 \begin{align}
 	\tau_{\gamma e,\rm C}=\frac{m_e}{4 n_e \sigma_T T_e}\approx 1.3 \times 10^{29}\left(\frac{T}{T_e}\right)(1+z)^{-4}\, \rm s\,.
 \end{align}
 This timescale is equal to that defined earlier in Ref.~\cite{Danese1977}.

  \textbf{Inverse Compton scattering}.--- The inverse Compton scattering of hot electrons on background photons is much faster than Compton scattering, since the background photon abundance is much larger than background electrons.   
 The time scale for   inverse Compton scattering, or the Compton cooling time of hot electrons, is given by~\cite{Danese1977}
 \begin{align}\label{eq:tau_invC}
 	\tau_{e\gamma, \rm invC}=\frac{3m_e}{4\sigma_T \rho_\gamma}\approx 7.4 \times 10^{19}(1+z)^{-4}~\rm s\,.
 \end{align}
 This can also be estimated by noting that in   kinetic equilibrium, the  loss rate for the thermally averaged   kinetic energy of electrons: $3 T_e n_e/(2	\tau_{e\gamma, \rm invC})$,  is equal to that for  the photon energy: $\rho_\gamma/ 	\tau_{\gamma e,\rm C}$, which leads to the timescale a factor of 2 smaller  than  Eq.~\eqref{eq:tau_invC}.

 \textbf{Double Compton scattering}.---
 The double Compton scattering rate can be written as
 \begin{align}\label{eq:dC}
 	\frac{\delta f}{\delta t}\Big|_{\text{dC}}&=n_e\sigma_T \frac{4\alpha_{\rm EM}}{3\pi}\left(\frac{T_e}{m_e}\right)^2 g_{\rm dC} \frac{e^{-x_e}}{x_e^3}\left[1-(e^{x_e}-1)f_\gamma\right],
 \end{align}
 where $g_{\rm dC}$ is the Gaunt factor. In the approximation of small photon spectrum distortions and small temperature difference $T_e-T$, the double Compton Gaunt factor can be fitted as~\cite{Chluba:2011hw}
 \begin{align}\label{eq:gdC}
 	g_{\rm dC}\approx \frac{I_{\rm dC}}{1+14.16 T/m_e}e^{-x_e}\left(1+\frac{3}{2}x_e+\frac{29}{24}x_e^2+\frac{11}{16}x_e^3+\frac{5}{12}x_e^4\right),
 \end{align}
with $I_{\rm dC}=4\pi^4/15$.  $g_{\rm dC}/I_{\rm dC}\approx 1$ for small $x_e\lesssim 1$ while $g_{\rm dC}/I_{\rm dC}$ is exponentially suppressed in the high-$x_e$ band.   Note that we have factored out a factor $e^{-x_e}$ from  the above result in Eq.~\eqref{eq:dC}. 
  
  The time scale of double Compton scattering  not only depends on the redshift but also on photon frequency.  The $x_e$-independent factor in Eq.~\eqref{eq:dC} characterizes the typical $z$-dependent timescale while   the efficiency of photon chemical thermalization in light of   the photon frequency is governed by the Gaunt factor.  To characterize the timescale for double Compton scattering that brings the perturbed photon distribution back to equilibrium, we can substitute $f_\gamma=(e^{x_e}-1)^{-1}+\delta f$ into Eq.~\eqref{eq:dC}. It leads to an  exponentially decreased solution to $\delta f$, with the damping timescale, or equivalently the thermalization timescale for double Compton scattering:
  \begin{align}\label{eq:tdC}
  	\tau_{\rm dC}&=\frac{3\pi}{4\alpha_{\rm EM} n_e\sigma_T }\left(\frac{m_e}{T_e}\right)^2\frac{x_e^3e^{x_e}}{g_{\rm dC}(e^{x_e}-1)}
  	\\[0.2cm]
  	&\approx 3.13\times 10^{41}\left(\frac{T}{T_e}\right)^2\frac{x_e^3e^{x_e}}{g_{\rm dC}(e^{x_e}-1)}(1+z)^{-5}\,\rm s\,,
  \end{align}
  where $n_e=(1-Y_p/2)n_B$ with the helium mass  fraction $Y_p\approx 0.243$ and baryon density $\Omega_B h^2\approx 0.0224$~\cite{Planck:2018vyg}. The photon temperature $T=T_0 (1+z)$ is introduced with $T_0=2.73$~K fixed.  For $x_e\gg 1$, $\tau_{\rm dC}$ increases exponentially, i.e., the  double Compton scattering rate is exponentially suppressed, while for $x_e\ll 1$, the timescale decreases quadratically. 
 
\textbf{Bremsstrahlung}.---The bremsstrahlung process carries a similar dependence on the photon distribution function, as given by~\cite{Hu:1992dc}
 \begin{align}\label{eq:Brem}
	\frac{\delta f}{\delta t}\Big|_{\text{Brem}}&= \frac{n_e\sigma_T}{m_e^3}2\sqrt{2\pi} \alpha_{\rm EM}\left(\frac{T_e}{m_e}\right)^{-7/2}\sum_i Z_i^2\, n_i\, g_{\rm Brem} \frac{e^{-x_e}}{x_e^3}\left[1-(e^{x_e}-1)f_\gamma\right],
\end{align}
 where $Z_i$ and $n_i$ denote the charge and number density for a nucleon species $i$, respectively. The bremsstrahlung Gaunt factor can be approximated as $g_{\rm Brem}\approx \ln(2.25/x_e)$ for $x_e<0.37$ and $g_{\rm Brem}\approx \pi/\sqrt{3}$ otherwise~\cite{Hu:1992dc}. Note that the
 small difference between $T_e$ and $T$ is neglected in the above rate. The rates from double Compton scattering and bremsstrahlung   imply that the chemical thermalization is still efficient in the infrared regime $x_e\to 0$ but becomes exponentially suppressed at the high-energy band $x_e\gg 1$, underlying  a frequency-dependent $\mu$-distortion. 
 
Similar to  Eq.~\eqref{eq:tdC}, the timescale for bremsstrahlung can be parameterized by
 \begin{align}
 	\tau_{\rm Brem}\approx8.36\times10^{26}\left(\frac{T_e}{T}\right)^{7/2}\frac{e^{x_e}x_e^3}{g_{\rm Brem}(e^{x_e}-1)}(1+z)^{-5/2}\, \rm s\,.
 	 \end{align}
For $x_e\gg 1$, $\tau_{\rm Brem}$ increases as $x_e^3$,   while for $x_e\ll 1$, the timescale decreases as $x_e^2 \ln x_e$.

\bibliographystyle{JHEP}
\bibliography{Refs}

\providecommand{\href}[2]{#2}\begingroup\raggedright\begin{thebibliography}{10}

\bibitem{Mather:1993ij}
J.~C. Mather et~al., {\it {Measurement of the Cosmic Microwave Background
  spectrum by the COBE FIRAS instrument}},  {\it Astrophys. J.} {\bf 420}
  (1994) 439--444.

\bibitem{Fixsen:1996nj}
D.~J. Fixsen, E.~S. Cheng, J.~M. Gales, J.~C. Mather, R.~A. Shafer, and E.~L.
  Wright, {\it {The Cosmic Microwave Background spectrum from the full COBE
  FIRAS data set}},  {\it Astrophys. J.} {\bf 473} (1996) 576,
  [\href{http://arxiv.org/abs/astro-ph/9605054}{{\tt astro-ph/9605054}}].

\bibitem{Fixsen_2009}
D.~J. Fixsen, {\it {The temperature of the cosmic microwave background}},  {\it
  ApJ.} {\bf 707} (2009) 916, [\href{http://arxiv.org/abs/0911.1955}{{\tt
  arXiv:0911.1955}}].

\bibitem{Burigana:1991eub}
C.~Burigana, L.~Danese, and G.~De~Zotti, {\it {Formation and evolution of early
  distortions of the microwave background spectrum - A numerical study}},  {\it
  Astron. Astrophys.} {\bf 246} (1991), no.~1 49--58.

\bibitem{Hu:1992dc}
W.~Hu and J.~Silk, {\it {Thermalization and spectral distortions of the cosmic
  background radiation}},  {\it Phys. Rev. D} {\bf 48} (1993) 485--502.

\bibitem{Chluba:2011hw}
J.~Chluba and R.~A. Sunyaev, {\it {The evolution of CMB spectral distortions in
  the early Universe}},  {\it Mon. Not. Roy. Astron. Soc.} {\bf 419} (2012)
  1294--1314, [\href{http://arxiv.org/abs/1109.6552}{{\tt arXiv:1109.6552}}].

\bibitem{Khatri:2012tw}
R.~Khatri and R.~A. Sunyaev, {\it {Beyond y and $\mu$: the shape of the CMB
  spectral distortions in the intermediate epoch, $1.5\times10^4 \lesssim z
  \lesssim 2\times 10^5$}},  {\it JCAP} {\bf 09} (2012) 016,
  [\href{http://arxiv.org/abs/1207.6654}{{\tt arXiv:1207.6654}}].

\bibitem{Acharya:2018iwh}
S.~K. Acharya and R.~Khatri, {\it {Rich structure of non-thermal relativistic
  CMB spectral distortions from high energy particle cascades at redshifts
  $z\lesssim 2\times 10^5$}},  {\it Phys. Rev. D} {\bf 99} (2019), no.~4
  043520, [\href{http://arxiv.org/abs/1808.02897}{{\tt arXiv:1808.02897}}].

\bibitem{Kogut:2011xw}
A.~Kogut et~al., {\it {The Primordial Inflation Explorer (PIXIE): A Nulling
  Polarimeter for Cosmic Microwave Background Observations}},  {\it JCAP} {\bf
  07} (2011) 025, [\href{http://arxiv.org/abs/1105.2044}{{\tt
  arXiv:1105.2044}}].

\bibitem{Chluba:2019nxa}
J.~Chluba et~al., {\it {New horizons in cosmology with spectral distortions of
  the cosmic microwave background}},  {\it Exper. Astron.} {\bf 51} (2021),
  no.~3 1515--1554, [\href{http://arxiv.org/abs/1909.01593}{{\tt
  arXiv:1909.01593}}].

\bibitem{Kogut:2019vqh}
A.~Kogut, M.~H. Abitbol, J.~Chluba, J.~Delabrouille, D.~Fixsen, J.~C. Hill,
  S.~P. Patil, and A.~Rotti, {\it {CMB Spectral Distortions: Status and
  Prospects}},  {\it Bull. Am. Astron. Soc.} {\bf 51} (2019), no.~7 113,
  [\href{http://arxiv.org/abs/1907.13195}{{\tt arXiv:1907.13195}}].

\bibitem{PRISM:2013fvg}
{\bf PRISM} Collaboration, P.~Andr\'e et~al., {\it {PRISM (Polarized Radiation
  Imaging and Spectroscopy Mission): An Extended White Paper}},  {\it JCAP}
  {\bf 02} (2014) 006, [\href{http://arxiv.org/abs/1310.1554}{{\tt
  arXiv:1310.1554}}].

\bibitem{Cabass:2016giw}
G.~Cabass, A.~Melchiorri, and E.~Pajer, {\it {$\mu$ distortions or running: A
  guaranteed discovery from CMB spectrometry}},  {\it Phys. Rev. D} {\bf 93}
  (2016), no.~8 083515, [\href{http://arxiv.org/abs/1602.05578}{{\tt
  arXiv:1602.05578}}].

\bibitem{Chluba:2016bvg}
J.~Chluba, {\it {Which spectral distortions does $\Lambda$CDM actually
  predict?}},  {\it Mon. Not. Roy. Astron. Soc.} {\bf 460} (2016), no.~1
  227--239, [\href{http://arxiv.org/abs/1603.02496}{{\tt arXiv:1603.02496}}].

\bibitem{Refregier:2000xz}
A.~Refregier, E.~Komatsu, D.~N. Spergel, and U.-L. Pen, {\it {Power spectrum of
  the Sunyaev-Zel'dovich effect}},  {\it Phys. Rev. D} {\bf 61} (2000) 123001,
  [\href{http://arxiv.org/abs/astro-ph/9912180}{{\tt astro-ph/9912180}}].

\bibitem{Dolag:2015dta}
K.~Dolag, E.~Komatsu, and R.~Sunyaev, {\it {SZ effects in the Magneticum
  Pathfinder Simulation: Comparison with the Planck, SPT, and ACT results}},
  {\it Mon. Not. Roy. Astron. Soc.} {\bf 463} (2016), no.~2 1797--1811,
  [\href{http://arxiv.org/abs/1509.05134}{{\tt arXiv:1509.05134}}].

\bibitem{Hill:2015tqa}
J.~C. Hill, N.~Battaglia, J.~Chluba, S.~Ferraro, E.~Schaan, and D.~N. Spergel,
  {\it {Taking the Universe\textquoteright{}s Temperature with Spectral
  Distortions of the Cosmic Microwave Background}},  {\it Phys. Rev. Lett.}
  {\bf 115} (2015), no.~26 261301, [\href{http://arxiv.org/abs/1507.01583}{{\tt
  arXiv:1507.01583}}].

\bibitem{DeZotti:2015awh}
G.~De~Zotti, M.~Negrello, G.~Castex, A.~Lapi, and M.~Bonato, {\it {Another look
  at distortions of the Cosmic Microwave Background spectrum}},  {\it JCAP}
  {\bf 03} (2016) 047, [\href{http://arxiv.org/abs/1512.04816}{{\tt
  arXiv:1512.04816}}].

\bibitem{Padmanabhan:2005es}
N.~Padmanabhan and D.~P. Finkbeiner, {\it {Detecting dark matter annihilation
  with CMB polarization: Signatures and experimental prospects}},  {\it Phys.
  Rev. D} {\bf 72} (2005) 023508,
  [\href{http://arxiv.org/abs/astro-ph/0503486}{{\tt astro-ph/0503486}}].

\bibitem{Chluba:2009uv}
J.~Chluba, {\it {Could the Cosmological Recombination Spectrum Help Us
  Understand Annihilating Dark Matter?}},  {\it Mon. Not. Roy. Astron. Soc.}
  {\bf 402} (2010) 1195, [\href{http://arxiv.org/abs/0910.3663}{{\tt
  arXiv:0910.3663}}].

\bibitem{Chluba:2013wsa}
J.~Chluba, {\it {Distinguishing different scenarios of early energy release
  with spectral distortions of the cosmic microwave background}},  {\it Mon.
  Not. Roy. Astron. Soc.} {\bf 436} (2013) 2232--2243,
  [\href{http://arxiv.org/abs/1304.6121}{{\tt arXiv:1304.6121}}].

\bibitem{McDonald:2000bk}
P.~McDonald, R.~J. Scherrer, and T.~P. Walker, {\it {Cosmic microwave
  background constraint on residual annihilations of relic particles}},  {\it
  Phys. Rev. D} {\bf 63} (2001) 023001,
  [\href{http://arxiv.org/abs/astro-ph/0008134}{{\tt astro-ph/0008134}}].

\bibitem{Chluba:2013pya}
J.~Chluba and D.~Jeong, {\it {Teasing bits of information out of the CMB energy
  spectrum}},  {\it Mon. Not. Roy. Astron. Soc.} {\bf 438} (2014), no.~3
  2065--2082, [\href{http://arxiv.org/abs/1306.5751}{{\tt arXiv:1306.5751}}].

\bibitem{ZS1970}
Y.~B. Zeldovich and R.~A. Sunyaev, {\it {The interaction of matter and
  radiation in the hot model of the Universe, II}},  {\it Astrophysics and
  Space Science} {\bf 7} (1970) 20.

\bibitem{Novikov2006}
I.~D.~N. Pavel D.~Naselsky, Dmitry I.~Novikov, {\it The physics of the cosmic
  microwave background}.
\newblock Cambridge University Press, 2006.

\bibitem{ZS1969}
Y.~B. Zeldovich and R.~A. Sunyaev, {\it {The Interaction of Matter and
  Radiation in a Hot-Model Universe}},  {\it Astrophysics and Space Science}
  {\bf 4} (1969) 301.

\bibitem{Chluba:2013kua}
J.~Chluba, {\it {Refined approximations for the distortion visibility function
  and \ensuremath{\mu}-type spectral distortions}},  {\it Mon. Not. Roy.
  Astron. Soc.} {\bf 440} (2014), no.~3 2544--2563,
  [\href{http://arxiv.org/abs/1312.6030}{{\tt arXiv:1312.6030}}].

\bibitem{Serpico:2004nm}
P.~D. Serpico and G.~G. Raffelt, {\it {MeV-mass dark matter and primordial
  nucleosynthesis}},  {\it Phys. Rev. D} {\bf 70} (2004) 043526,
  [\href{http://arxiv.org/abs/astro-ph/0403417}{{\tt astro-ph/0403417}}].

\bibitem{Ho:2012ug}
C.~M. Ho and R.~J. Scherrer, {\it {Limits on MeV Dark Matter from the Effective
  Number of Neutrinos}},  {\it Phys. Rev. D} {\bf 87} (2013), no.~2 023505,
  [\href{http://arxiv.org/abs/1208.4347}{{\tt arXiv:1208.4347}}].

\bibitem{Boehm:2013jpa}
C.~Boehm, M.~J. Dolan, and C.~McCabe, {\it {A Lower Bound on the Mass of Cold
  Thermal Dark Matter from Planck}},  {\it JCAP} {\bf 08} (2013) 041,
  [\href{http://arxiv.org/abs/1303.6270}{{\tt arXiv:1303.6270}}].

\bibitem{Nollett:2013pwa}
K.~M. Nollett and G.~Steigman, {\it {BBN And The CMB Constrain Light,
  Electromagnetically Coupled WIMPs}},  {\it Phys. Rev. D} {\bf 89} (2014),
  no.~8 083508, [\href{http://arxiv.org/abs/1312.5725}{{\tt arXiv:1312.5725}}].

\bibitem{Nollett:2014lwa}
K.~M. Nollett and G.~Steigman, {\it {BBN And The CMB Constrain Neutrino Coupled
  Light WIMPs}},  {\it Phys. Rev. D} {\bf 91} (2015), no.~8 083505,
  [\href{http://arxiv.org/abs/1411.6005}{{\tt arXiv:1411.6005}}].

\bibitem{Kawasaki:2015yya}
M.~Kawasaki, K.~Kohri, T.~Moroi, and Y.~Takaesu, {\it {Revisiting Big-Bang
  Nucleosynthesis Constraints on Dark-Matter Annihilation}},  {\it Phys. Lett.
  B} {\bf 751} (2015) 246--250, [\href{http://arxiv.org/abs/1509.03665}{{\tt
  arXiv:1509.03665}}].

\bibitem{Depta:2019lbe}
P.~F. Depta, M.~Hufnagel, K.~Schmidt-Hoberg, and S.~Wild, {\it {BBN constraints
  on the annihilation of MeV-scale dark matter}},  {\it JCAP} {\bf 04} (2019)
  029, [\href{http://arxiv.org/abs/1901.06944}{{\tt arXiv:1901.06944}}].

\bibitem{Sabti:2019mhn}
N.~Sabti, J.~Alvey, M.~Escudero, M.~Fairbairn, and D.~Blas, {\it {Refined
  Bounds on MeV-scale Thermal Dark Sectors from BBN and the CMB}},  {\it JCAP}
  {\bf 01} (2020) 004, [\href{http://arxiv.org/abs/1910.01649}{{\tt
  arXiv:1910.01649}}].

\bibitem{Ballesteros:2020adh}
G.~Ballesteros, M.~A.~G. Garcia, and M.~Pierre, {\it {How warm are non-thermal
  relics? Lyman-$\alpha$ bounds on out-of-equilibrium dark matter}},  {\it
  JCAP} {\bf 03} (2021) 101, [\href{http://arxiv.org/abs/2011.13458}{{\tt
  arXiv:2011.13458}}].

\bibitem{DEramo:2020gpr}
F.~D'Eramo and A.~Lenoci, {\it {Lower mass bounds on FIMP dark matter produced
  via freeze-in}},  {\it JCAP} {\bf 10} (2021) 045,
  [\href{http://arxiv.org/abs/2012.01446}{{\tt arXiv:2012.01446}}].

\bibitem{Decant:2021mhj}
Q.~Decant, J.~Heisig, D.~C. Hooper, and L.~Lopez-Honorez, {\it
  {Lyman-\ensuremath{\alpha} constraints on freeze-in and superWIMPs}},  {\it
  JCAP} {\bf 03} (2022) 041, [\href{http://arxiv.org/abs/2111.09321}{{\tt
  arXiv:2111.09321}}].

\bibitem{Cyr:2024vkt}
B.~Cyr, S.~K. Acharya, and J.~Chluba, {\it {Soft Photon Heating: A
  Semi-Analytic Framework and Applications to $21$cm Cosmology}},
  \href{http://arxiv.org/abs/2404.11743}{{\tt arXiv:2404.11743}}.

\bibitem{Planck:2018vyg}
{\bf Planck} Collaboration, N.~Aghanim et~al., {\it {Planck 2018 results. VI.
  Cosmological parameters}},  {\it Astron. Astrophys.} {\bf 641} (2020) A6,
  [\href{http://arxiv.org/abs/1807.06209}{{\tt arXiv:1807.06209}}]. [Erratum:
  Astron.Astrophys. 652, C4 (2021)].

\bibitem{Kolb:1990vq}
E.~W. Kolb and M.~S. Turner, {\it {The Early Universe}}, vol.~69.
\newblock 1990.

\bibitem{Slatyer:2009yq}
T.~R. Slatyer, N.~Padmanabhan, and D.~P. Finkbeiner, {\it {CMB Constraints on
  WIMP Annihilation: Energy Absorption During the Recombination Epoch}},  {\it
  Phys. Rev. D} {\bf 80} (2009) 043526,
  [\href{http://arxiv.org/abs/0906.1197}{{\tt arXiv:0906.1197}}].

\bibitem{Sommerfeld:1931qaf}
A.~Sommerfeld, {\it On the diffraction and deceleration of electrons},  {\it
  Annalen Phys.} {\bf 403} (1931), no.~3 257--330.

\bibitem{Zavala:2009mi}
J.~Zavala, M.~Vogelsberger, and S.~D.~M. White, {\it {Relic density and CMB
  constraints on dark matter annihilation with Sommerfeld enhancement}},  {\it
  Phys. Rev. D} {\bf 81} (2010) 083502,
  [\href{http://arxiv.org/abs/0910.5221}{{\tt arXiv:0910.5221}}].

\bibitem{Arkani-Hamed:2008hhe}
N.~Arkani-Hamed, D.~P. Finkbeiner, T.~R. Slatyer, and N.~Weiner, {\it {A Theory
  of Dark Matter}},  {\it Phys. Rev. D} {\bf 79} (2009) 015014,
  [\href{http://arxiv.org/abs/0810.0713}{{\tt arXiv:0810.0713}}].

\bibitem{Feng:2009hw}
J.~L. Feng, M.~Kaplinghat, and H.-B. Yu, {\it {Halo Shape and Relic Density
  Exclusions of Sommerfeld-Enhanced Dark Matter Explanations of Cosmic Ray
  Excesses}},  {\it Phys. Rev. Lett.} {\bf 104} (2010) 151301,
  [\href{http://arxiv.org/abs/0911.0422}{{\tt arXiv:0911.0422}}].

\bibitem{Feng:2010zp}
J.~L. Feng, M.~Kaplinghat, and H.-B. Yu, {\it {Sommerfeld Enhancements for
  Thermal Relic Dark Matter}},  {\it Phys. Rev. D} {\bf 82} (2010) 083525,
  [\href{http://arxiv.org/abs/1005.4678}{{\tt arXiv:1005.4678}}].

\bibitem{Bringmann:2006mu}
T.~Bringmann and S.~Hofmann, {\it {Thermal decoupling of WIMPs from first
  principles}},  {\it JCAP} {\bf 04} (2007) 016,
  [\href{http://arxiv.org/abs/hep-ph/0612238}{{\tt hep-ph/0612238}}]. [Erratum:
  JCAP 03, E02 (2016)].

\bibitem{Green:2018pmd}
D.~Green, P.~D. Meerburg, and J.~Meyers, {\it {Aspects of Dark Matter
  Annihilation in Cosmology}},  {\it JCAP} {\bf 04} (2019) 025,
  [\href{http://arxiv.org/abs/1804.01055}{{\tt arXiv:1804.01055}}].

\bibitem{Cang:2020exa}
J.~Cang, Y.~Gao, and Y.-Z. Ma, {\it {Probing dark matter with future CMB
  measurements}},  {\it Phys. Rev. D} {\bf 102} (2020), no.~10 103005,
  [\href{http://arxiv.org/abs/2002.03380}{{\tt arXiv:2002.03380}}].

\bibitem{Cyr:2023pgw}
B.~Cyr, T.~Kite, J.~Chluba, J.~C. Hill, D.~Jeong, S.~K. Acharya, B.~Bolliet,
  and S.~P. Patil, {\it {Disentangling the primordial nature of stochastic
  gravitational wave backgrounds with CMB spectral distortions}},  {\it Mon.
  Not. Roy. Astron. Soc.} {\bf 528} (2024), no.~1 883--897,
  [\href{http://arxiv.org/abs/2309.02366}{{\tt arXiv:2309.02366}}].

\bibitem{McDonald:2001vt}
J.~McDonald, {\it {Thermally generated gauge singlet scalars as selfinteracting
  dark matter}},  {\it Phys. Rev. Lett.} {\bf 88} (2002) 091304,
  [\href{http://arxiv.org/abs/hep-ph/0106249}{{\tt hep-ph/0106249}}].

\bibitem{Kusenko:2006rh}
A.~Kusenko, {\it {Sterile neutrinos, dark matter, and the pulsar velocities in
  models with a Higgs singlet}},  {\it Phys. Rev. Lett.} {\bf 97} (2006)
  241301, [\href{http://arxiv.org/abs/hep-ph/0609081}{{\tt hep-ph/0609081}}].

\bibitem{Petraki:2007gq}
K.~Petraki and A.~Kusenko, {\it {Dark-matter sterile neutrinos in models with a
  gauge singlet in the Higgs sector}},  {\it Phys. Rev. D} {\bf 77} (2008)
  065014, [\href{http://arxiv.org/abs/0711.4646}{{\tt arXiv:0711.4646}}].

\bibitem{Hall:2009bx}
L.~J. Hall, K.~Jedamzik, J.~March-Russell, and S.~M. West, {\it {Freeze-In
  Production of FIMP Dark Matter}},  {\it JHEP} {\bf 03} (2010) 080,
  [\href{http://arxiv.org/abs/0911.1120}{{\tt arXiv:0911.1120}}].

\bibitem{Bernal:2017kxu}
N.~Bernal, M.~Heikinheimo, T.~Tenkanen, K.~Tuominen, and V.~Vaskonen, {\it {The
  Dawn of FIMP Dark Matter: A Review of Models and Constraints}},  {\it Int. J.
  Mod. Phys. A} {\bf 32} (2017), no.~27 1730023,
  [\href{http://arxiv.org/abs/1706.07442}{{\tt arXiv:1706.07442}}].

\bibitem{Pospelov:2007mp}
M.~Pospelov, A.~Ritz, and M.~B. Voloshin, {\it {Secluded WIMP Dark Matter}},
  {\it Phys. Lett. B} {\bf 662} (2008) 53--61,
  [\href{http://arxiv.org/abs/0711.4866}{{\tt arXiv:0711.4866}}].

\bibitem{Evans:2017kti}
J.~A. Evans, S.~Gori, and J.~Shelton, {\it {Looking for the WIMP Next Door}},
  {\it JHEP} {\bf 02} (2018) 100, [\href{http://arxiv.org/abs/1712.03974}{{\tt
  arXiv:1712.03974}}].

\bibitem{Kanemura:2023jiw}
S.~Kanemura and S.-P. Li, {\it {Dark phase transition from WIMP: complementary
  tests from gravitational waves and colliders}},
  \href{http://arxiv.org/abs/2308.16390}{{\tt arXiv:2308.16390}}.

\bibitem{Sabti:2021reh}
N.~Sabti, J.~Alvey, M.~Escudero, M.~Fairbairn, and D.~Blas, {\it {Addendum:
  Refined bounds on MeV-scale thermal dark sectors from BBN and the CMB}},
  {\it JCAP} {\bf 08} (2021) A01, [\href{http://arxiv.org/abs/2107.11232}{{\tt
  arXiv:2107.11232}}].

\bibitem{Berlin:2017ftj}
A.~Berlin and N.~Blinov, {\it {Thermal Dark Matter Below an MeV}},  {\it Phys.
  Rev. Lett.} {\bf 120} (2018), no.~2 021801,
  [\href{http://arxiv.org/abs/1706.07046}{{\tt arXiv:1706.07046}}].

\bibitem{Reno:1987qw}
M.~H. Reno and D.~Seckel, {\it {Primordial Nucleosynthesis: The Effects of
  Injecting Hadrons}},  {\it Phys. Rev. D} {\bf 37} (1988) 3441.

\bibitem{Green:2005fa}
A.~M. Green, S.~Hofmann, and D.~J. Schwarz, {\it {The First wimpy halos}},
  {\it JCAP} {\bf 08} (2005) 003,
  [\href{http://arxiv.org/abs/astro-ph/0503387}{{\tt astro-ph/0503387}}].

\bibitem{Vogelsberger:2015gpr}
M.~Vogelsberger, J.~Zavala, F.-Y. Cyr-Racine, C.~Pfrommer, T.~Bringmann, and
  K.~Sigurdson, {\it {ETHOS \textendash{} an effective theory of structure
  formation: dark matter physics as a possible explanation of the small-scale
  CDM problems}},  {\it Mon. Not. Roy. Astron. Soc.} {\bf 460} (2016), no.~2
  1399--1416, [\href{http://arxiv.org/abs/1512.05349}{{\tt arXiv:1512.05349}}].

\bibitem{vandenAarssen:2012vpm}
L.~G. van~den Aarssen, T.~Bringmann, and C.~Pfrommer, {\it {Is dark matter with
  long-range interactions a solution to all small-scale problems of
  \textbackslash{}Lambda CDM cosmology?}},  {\it Phys. Rev. Lett.} {\bf 109}
  (2012) 231301, [\href{http://arxiv.org/abs/1205.5809}{{\tt
  arXiv:1205.5809}}].

\bibitem{Chu:2014lja}
X.~Chu and B.~Dasgupta, {\it {Dark Radiation Alleviates Problems with Dark
  Matter Halos}},  {\it Phys. Rev. Lett.} {\bf 113} (2014), no.~16 161301,
  [\href{http://arxiv.org/abs/1404.6127}{{\tt arXiv:1404.6127}}].

\bibitem{Bringmann:2016ilk}
T.~Bringmann, H.~T. Ihle, J.~Kersten, and P.~Walia, {\it {Suppressing structure
  formation at dwarf galaxy scales and below: late kinetic decoupling as a
  compelling alternative to warm dark matter}},  {\it Phys. Rev. D} {\bf 94}
  (2016), no.~10 103529, [\href{http://arxiv.org/abs/1603.04884}{{\tt
  arXiv:1603.04884}}].

\bibitem{Tulin:2017ara}
S.~Tulin and H.-B. Yu, {\it {Dark Matter Self-interactions and Small Scale
  Structure}},  {\it Phys. Rept.} {\bf 730} (2018) 1--57,
  [\href{http://arxiv.org/abs/1705.02358}{{\tt arXiv:1705.02358}}].

\bibitem{Bullock:2017xww}
J.~S. Bullock and M.~Boylan-Kolchin, {\it {Small-Scale Challenges to the
  $\Lambda$CDM Paradigm}},  {\it Ann. Rev. Astron. Astrophys.} {\bf 55} (2017)
  343--387, [\href{http://arxiv.org/abs/1707.04256}{{\tt arXiv:1707.04256}}].

\bibitem{Benito:2020avv}
M.~Benito, J.~C. Criado, G.~H\"utsi, M.~Raidal, and H.~Veerm\"ae, {\it
  {Implications of Milky Way substructures for the nature of dark matter}},
  {\it Phys. Rev. D} {\bf 101} (2020), no.~10 103023,
  [\href{http://arxiv.org/abs/2001.11013}{{\tt arXiv:2001.11013}}].

\bibitem{Irsic:2017ixq}
V.~Ir\v{s}i\v{c} et~al., {\it {New Constraints on the free-streaming of warm
  dark matter from intermediate and small scale Lyman-$\alpha$ forest data}},
  {\it Phys. Rev. D} {\bf 96} (2017), no.~2 023522,
  [\href{http://arxiv.org/abs/1702.01764}{{\tt arXiv:1702.01764}}].

\bibitem{DES:2020fxi}
{\bf DES} Collaboration, E.~O. Nadler et~al., {\it {Milky Way Satellite Census.
  III. Constraints on Dark Matter Properties from Observations of Milky Way
  Satellite Galaxies}},  {\it Phys. Rev. Lett.} {\bf 126} (2021) 091101,
  [\href{http://arxiv.org/abs/2008.00022}{{\tt arXiv:2008.00022}}].

\bibitem{Wilkinson:2013kia}
R.~J. Wilkinson, J.~Lesgourgues, and C.~Boehm, {\it {Using the CMB angular
  power spectrum to study Dark Matter-photon interactions}},  {\it JCAP} {\bf
  04} (2014) 026, [\href{http://arxiv.org/abs/1309.7588}{{\tt
  arXiv:1309.7588}}].

\bibitem{Boudaud:2018oya}
M.~Boudaud, T.~Lacroix, M.~Stref, and J.~Lavalle, {\it {Robust cosmic-ray
  constraints on $p$-wave annihilating MeV dark matter}},  {\it Phys. Rev. D}
  {\bf 99} (2019), no.~6 061302, [\href{http://arxiv.org/abs/1810.01680}{{\tt
  arXiv:1810.01680}}].

\bibitem{Genolini:2021doh}
Y.~G\'enolini, M.~Boudaud, M.~Cirelli, L.~Derome, J.~Lavalle, D.~Maurin,
  P.~Salati, and N.~Weinrich, {\it {New minimal, median, and maximal
  propagation models for dark matter searches with Galactic cosmic rays}},
  {\it Phys. Rev. D} {\bf 104} (2021), no.~8 083005,
  [\href{http://arxiv.org/abs/2103.04108}{{\tt arXiv:2103.04108}}].

\bibitem{Genolini:2019ewc}
Y.~G\'enolini et~al., {\it {Cosmic-ray transport from AMS-02 boron to carbon
  ratio data: Benchmark models and interpretation}},  {\it Phys. Rev. D} {\bf
  99} (2019), no.~12 123028, [\href{http://arxiv.org/abs/1904.08917}{{\tt
  arXiv:1904.08917}}].

\bibitem{Boddy:2015efa}
K.~K. Boddy and J.~Kumar, {\it {Indirect Detection of Dark Matter Using
  MeV-Range Gamma-Ray Telescopes}},  {\it Phys. Rev. D} {\bf 92} (2015), no.~2
  023533, [\href{http://arxiv.org/abs/1504.04024}{{\tt arXiv:1504.04024}}].

\bibitem{Albert:2014hwa}
{\bf Fermi-LAT} Collaboration, A.~Albert, G.~A. Gomez-Vargas, M.~Grefe,
  C.~Munoz, C.~Weniger, E.~D. Bloom, E.~Charles, M.~N. Mazziotta, and
  A.~Morselli, {\it {Search for 100 MeV to 10 GeV $\gamma$-ray lines in the
  Fermi-LAT data and implications for gravitino dark matter in $\mu\nu$SSM}},
  {\it JCAP} {\bf 10} (2014) 023, [\href{http://arxiv.org/abs/1406.3430}{{\tt
  arXiv:1406.3430}}].

\bibitem{Essig:2013goa}
R.~Essig, E.~Kuflik, S.~D. McDermott, T.~Volansky, and K.~M. Zurek, {\it
  {Constraining Light Dark Matter with Diffuse X-Ray and Gamma-Ray
  Observations}},  {\it JHEP} {\bf 11} (2013) 193,
  [\href{http://arxiv.org/abs/1309.4091}{{\tt arXiv:1309.4091}}].

\bibitem{Chluba:2015hma}
J.~Chluba, {\it {Green's function of the cosmological thermalization problem
  \textendash{} II. Effect of photon injection and constraints}},  {\it Mon.
  Not. Roy. Astron. Soc.} {\bf 454} (2015), no.~4 4182--4196,
  [\href{http://arxiv.org/abs/1506.06582}{{\tt arXiv:1506.06582}}].

\bibitem{Sazonov:1999fy}
S.~Y. Sazonov and R.~A. Sunyaev, {\it {The Profile of a narrow line after
  single scattering by maxwellian electrons: Relativistic corrections to the
  kernel of the integral kinetic equation}},  {\it Astrophys. J.} {\bf 543}
  (2000) 28--55, [\href{http://arxiv.org/abs/astro-ph/9910280}{{\tt
  astro-ph/9910280}}].

\bibitem{Danese1977}
L.~Danese and G.~De~Zotti, {\it {The relic radiation spectrum and the thermal
  history of the universe}},  {\it Riv. Nuovo Cim.} {\bf 7} (1977) 277--362.

\end{thebibliography}\endgroup

\end{document}